\pdfoutput=1
\documentclass[cernpreprint, atlasdraft=false, UKenglish, texmf, orcidlogo]{atlasdoc}
\usepackage[subcaption]{atlaspackage}
\usepackage{atlasbiblatex}

\usepackage[jetetmiss, BSM=true, misc=true]{atlasphysics}

\usepackage{siunitx}

\graphicspath{{logos/}{figures/}}

\usepackage{ANA-SUSY-2021-07-PAPER-defs}

\AtlasTitle{Search for signatures of electroweakinos with photons, jets, and large missing transverse momentum in $\sqrt{s}=13$~\TeV $pp$ collisions with the ATLAS detector}

\AtlasAbstract{%
A search for final states characterised by at least one isolated high transverse-momentum photon, jets and large missing transverse momentum is presented. Such a final state might occur in gauge-mediated supersymmetric models where a pair of binos and higgsinos mix to form neutralinos, one of which decays into a photon plus a gravitino while the other decays into a Higgs boson, a $Z$ boson or a photon, plus a gravitino. The search is performed using the full Run-2 data sample of 140 fb$^{-1}$ of $\sqrt{s}=13$ TeV proton--proton collisions collected by the ATLAS detector at the Large Hadron Collider. No significant excess of events is observed above the Standard Model prediction and model-dependent exclusion limits at the 95\% confidence level are set.
These limits are interpreted in terms of the masses of gauginos, which are excluded up to 1.2 TeV depending on their branching ratios, and on their branching ratios as a function of their mass.
}

\PreprintIdNumber{CERN-EP-2025-198}

\arXivId{2511.21240}

\HepDataRecord{168319}

\AtlasJournalRef{\JHEP 04 (2026) 150}
\AtlasDOI{10.1007/JHEP04(2026)150}

\hypersetup{pdftitle={ATLAS document},pdfauthor={The ATLAS Collaboration}}

\begin{document}

\maketitle

\tableofcontents

\section{Introduction}
%
%

This paper describes a search for an excess of events with isolated high transverse-momentum (high-\pt) photons, jets and large missing transverse momentum (\ETM, with magnitude \met) over the Standard Model (SM) prediction that may constitute a striking signature of physics beyond the SM.

The results are interpreted in the context of supersymmetry (SUSY)~\cite{Golfand:1971iw,Volkov:1973ix,Wess:1974tw,Wess:1974jb,Ferrara:1974pu,Salam:1974ig,Martin:1997ns}, a theoretical extension of the SM that solves the hierarchy problem \cite{Barbieri:1987fn,deCarlos:1993yy} through the addition of a new fermionic/bosonic supersymmetric partner to each boson/fermion in the SM\@ together with an extended Higgs sector with two Higgs doublets. Theories of gauge-mediated SUSY breaking (GMSB)~\cite{Dine:1981gu,AlvarezGaume:1981wy,Nappi:1982hm} presume a hidden sector in which supersymmetry is broken by a vacuum expectation value, and the symmetry breaking is communicated to the visible sector through SM gauge boson interactions. In SUSY models with low-energy SUSY breaking scales, such as GMSB or general gauge mediation (GGM)~\cite{Cheung:2007es,Meade:2008wd}, the ultra-light gravitino (\gravino) is the lightest supersymmetric particle (LSP) and stable under R-parity conservation. Therefore, the gravitino is a candidate for dark matter~\cite{Goldberg:1983nd,Ellis:1983ew}. In this paper, GGM models are considered, which are extensions of the GMSB idea, allowing decoupled mass scales for strongly interacting supersymmetric partners.

A general feature of SUSY models is the mixing of binos, winos and higgsinos (the superpartners of the electroweak gauge bosons and the Higgs bosons, respectively). These electroweakinos mix to form neutralino ($\ninoone,\ninotwo, \ninothree,\ninofour$) and chargino ($\chinotwopm,\chinoonepm$) mass eigenstates, where the states are ordered by increasing values of their mass.

The phenomenology of the models is driven by the nature of the next-to-lightest SUSY particle (NLSP). In this study, the NLSP is assumed to be the lightest neutralino {\ninoone}, as occurs in a large fraction of the GGM parameter space.
The {\ninoone} decays into a \gravino\ and either a $\gam$, a $Z$ boson, or the lightest neutral Higgs boson $h$, assumed to be the 125~\GeV Higgs boson observed by ATLAS~\cite{HIGG-2012-27} and CMS~\cite{CMS-HIG-12-028}.

If the {\ninoone} is predominantly bino, the main decay channel is $\ninoone\to\gam\gravino$, while a state with a dominant higgsino component decays as $\ninoone\to h\gravino$. In addition, since the longitudinal polarisation component of the $Z$ boson is also a Goldstone mode of the Higgs field, a higgsino admixture neutralino can also decay as $\ninoone\to Z\gravino$.

Consequently, a pair of {\ninoone} produced in a hadron collider can give rise to a signature containing two final-state bosons ($hh$, $h\gam$, $hZ$, $Z\gam$, $ZZ$ or $\gamma\gamma$) plus \etmiss\ from the stable, undetectable LSP particles (\gravino). Several analyses using these signatures have been conducted in the search for SUSY particles produced via strong interactions at the LHC~\cite{SUSY-2011-12,SUSY-2012-12,EXOT-2010-06,SUSY-2011-04,CMS-SUS-10-002,CMS-SUS-11-002,CMS-SUS-14-009,SUSY-2014-01,CMS-SUS-14-004,SUSY-2014-08,SUSY-2016-27,SUSY-2018-11}.
A previous search by ATLAS for the case of electroweak production was performed using data from $pp$ collisions at $\sqrt{s}=8\ \tev$ corresponding to an integrated luminosity of 20.3 \invfb~\cite{SUSY-2014-01}.
In Run~2 of the Large Hadron Collider (LHC)~\cite{Evans:2008zzb}, previous ATLAS searches for electroweakinos have focused on exclusive diphoton signatures, initially using partial data~\cite{SUSY-2016-27}, and later with the full data sample, targeting Higgs boson decays into two photons~\cite{SUSY-2018-23,SUSY-2020-17}. More recently, CMS has reported their corresponding Run-2 results using a similar signature and targeting related models~\cite{CMS-SUS-21-009}.

The present result is based on the full Run-2 ATLAS data sample and complements earlier searches focused on strong production channels. This analysis enhances the sensitivity to electroweak production by probing new regions of the parameter space, employing a more inclusive selection of final states, and extending the reach to higher masses than previous ATLAS electroweakinos searches, investigating neutralino masses up to 1.2~\TeV. Another key innovation in this study is the scanning over different combinations of branching ratios to derive exclusion limits, significantly broadening the scope compared to previous analyses.

\begin{figure}[ht!]
\centering
\includegraphics[width=0.45\textwidth]{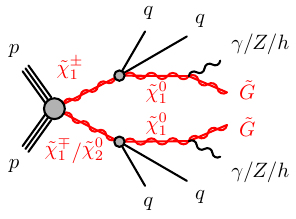}
\includegraphics[width=0.45\textwidth]{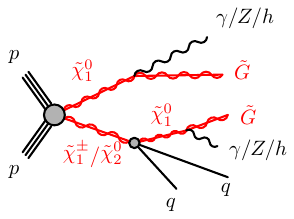}

\caption{Example production diagrams of neutralinos and their subsequent decay into a final state with a photon, jets and missing transverse momentum.}
\label{fig:GGM_diagrams}
\end{figure}

The analysis presented in this paper targets the production of {\ninoone} considered as a bino--higgsino admixture resulting in final states with at least one \gam plus two \gravino. Example diagrams of the processes in the R-parity conservation GGM model are shown in Figure~\ref{fig:GGM_diagrams}.
The phenomenology of bino–higgsino models and their implications for LHC searches are also discussed in the context of extended Higgs sector scenarios in Ref. \cite{Liu:2020muv}.

%
\section{ATLAS detector}
%
%

%
\newcommand{\AtlasCoordFootnote}{%
ATLAS uses a right-handed coordinate system with its origin at the nominal interaction point (IP)
in the centre of the detector and the \(z\)-axis along the beam pipe.
The \(x\)-axis points from the IP to the centre of the LHC ring,
and the \(y\)-axis points upwards.
Polar coordinates \((r,\phi)\) are used in the transverse plane,
\(\phi\) being the azimuthal angle around the \(z\)-axis.
The pseudorapidity is defined in terms of the polar angle \(\theta\) as \(\eta = -\ln \tan(\theta/2)\) and is equal to the rapidity
$ y = \frac{1}{2} \ln \left( \frac{E + p_z}{E - p_z} \right) $ in the relativistic limit.
Angular distance is measured in units of \(\Delta R \equiv \sqrt{(\Delta y)^{2} + (\Delta\phi)^{2}}\).}

The ATLAS detector~\cite{PERF-2007-01} at the LHC covers nearly the entire solid angle around the collision point.\footnote{\AtlasCoordFootnote}
It consists of an inner tracking detector surrounded by a thin superconducting solenoid, electromagnetic (EM) and hadron calorimeters,
and a muon spectrometer incorporating three large superconducting air-core toroidal magnets.

The inner-detector system (ID) is immersed in a \qty{2}{\tesla} axial magnetic field
and provides charged-particle tracking in the range of \(|\eta| < 2.5\).
The high-granularity silicon pixel detector covers the vertex region and typically provides four measurements per track,
the first hit normally being in the insertable B-layer (IBL) installed before Run~2~\cite{ATLAS-TDR-19,PIX-2018-001}.
It is followed by the silicon microstrip tracker (SCT), which usually provides eight measurements per track.
These silicon detectors are complemented by the transition radiation tracker (TRT),
which enables radially extended track reconstruction up to \(|\eta| = 2.0\).
The TRT also provides electron identification information
based on the fraction of hits (typically 30 in total) above a higher energy-deposit threshold corresponding to transition radiation.

The calorimeter system covers the pseudorapidity range \(|\eta| < 4.9\).
Within the region \(|\eta|< 3.2\), EM calorimetry is provided by barrel and
endcap high-granularity lead/liquid-argon (LAr) calorimeters,
with an additional thin LAr presampler covering \(|\eta| < 1.8\)
to correct for energy loss in material upstream of the calorimeters.
Hadron calorimetry is provided by a steel/scintillator-tile calorimeter,
segmented into three barrel structures within \(|\eta| < 1.7\), and two copper/LAr hadron endcap calorimeters.
The solid angle coverage is completed with forward copper/LAr and tungsten/LAr calorimeter modules
optimised for EM and hadronic energy measurements respectively.

The muon spectrometer (MS) comprises separate trigger and
high-precision tracking chambers measuring the deflection of muons in a magnetic field generated by the superconducting air-core toroidal magnets.
The field integral of the toroids ranges between \num{2.0} and \qty{6.0}{\tesla\metre}
across most of the detector.
Three layers of precision chambers, each consisting of layers of monitored drift tubes, cover the region \(|\eta| < 2.7\),
complemented by cathode-strip chambers in the forward region, where the background is highest.
The muon trigger system covers the range \(|\eta| < 2.4\) with resistive-plate chambers in the barrel, and thin-gap chambers in the endcap regions.

Interesting events are selected by the first-level trigger system implemented in custom hardware,
followed by selections made by algorithms implemented in software in the high-level trigger~\cite{TRIG-2016-01}.
The first-level trigger accepts events from the \qty{40}{\MHz} bunch crossings at a rate close to \qty{100}{\kHz},
which the high-level trigger further reduces in order to record complete events to disk at about \qty{1.25}{\kHz}.

A software suite~\cite{SOFT-2022-02} is used in data simulation, in the reconstruction and analysis of real and simulated data, in detector operations, and in the trigger and data acquisition systems of the experiment.


%
\section{Simulation samples}
%
%

Samples of the targeted GGM SUSY signals and SM backgrounds were simulated at \ECMS\ using dedicated Monte Carlo (MC) generators. For the design of the signal regions (SR) and the interpretation of the results a collection of signal samples was generated selecting a set of benchmark points.
The bino component of the lightest neutralino couples to both the photon and the $Z$ boson, while the higgsino component couples to the $Z$ or Higgs boson. To generate the signal samples a simplified approach was implemented, where the masses of the particles, the decay branching fractions and the mixing matrices were the only input parameters.
The four most relevant electroweakino production channels were considered: \ninoone\ninotwo, \ninoone\chinoonepm, \ninotwo\chinoonepm and \chinoonep\chinoonem. Production modes including \ninothree are highly suppressed and are not considered for this search. The samples corresponding to the signal points were produced as a function of the {\ninoone} mass, $m_{\ninoone}$, from 150 to 1450~\gev in steps of 100~\gev.
The {\ninotwo} (\chinoonepm) mass was chosen to be equal to $m_{\ninoone} + 11 (10) \GeV$, while the {\gravino} mass was set to 1~\eV. %
The rest of the sparticle masses were set in the decoupled regime at 4.5~\TeV. The {\ninotwo} (\chinoonepm) were set to always decay into {\ninoone} via an off-shell $Z$ ($W^{\pm}$), with the latter having SM decays. The {\ninoone} branching fractions were set to be equal between $\gravino+\gam$, $\gravino+Z$ and $\gravino+h$ decays. This choice allows for a re-weighting of the events in order to reproduce different models, characterised by the {\ninoone} branching ratio, and also offers the possibility to set exclusion limits as a function of the branching ratio using only one set of signal samples. All signal events contain a pair of {\ninoone} in the decay chain, so each event can be classified by the decay product of the pair, i.e. $\gam\gam, \gam h, \gam Z, ZZ, Zh$ or $hh$. Two benchmark scenarios are studied as the most representative for the final state with photons and jets: the {\ninoone} decaying 50\% into $\gam+\gravino$ and 50\% into $Z+\gravino$, and the {\ninoone} decaying 50\% into $\gam+\gravino$ and 50\% into $h+\gravino$ (referred to as the `$\gam+Z$' and `$\gam+h$' benchmark models respectively). All signal samples were generated at leading order (LO) in quantum chromodynamics (QCD) with up to two additional partons with \MADGRAPH[5] ~\cite{Alwall:2014hca} interfaced to \PYTHIA[8.2] ~\cite{Sjostrand:2014zea} using the A14 tune and the \NNPDF[2.3lo] PDF set\cite{Ball:2012cx}.

The cross-section of each process was calculated using \textsc{resummino-3.0.0} \cite{Beenakker:1999xh,Debove:2010kf,Fuks:2012qx,Fuks:2013vua,Fiaschi:2018hgm} at next-to-leading plus next-to-leading-logarithmic (NLO+NLL) precision, using CTEQ6.6 and the MSTW2008 parton distribution functions (PDFs),  with their corresponding variations in PDF sets and in scales as described in Ref. \cite{Borschensky:2014cia}.

Samples of \ttbargam\ events were generated with \MGNLO~\cite{Alwall:2014hca} at NLO (with $m_{\text{top}}=172.5~\gev$),
interfaced to the \PYTHIA[8.2] parton shower model~\cite{Sjostrand:2014zea} with the A14 set of tuned parameters (tune)~\cite{ATL-PHYS-PUB-2014-021}. The \NNPDF[3.0nlo]~\cite{Ball:2014uwa} set of PDFs was used.  The production of Higgs boson events, {\tthgamgam}, was modelled using the  \POWHEGBOX[v2]~\cite{Baernreuther:2012ws,Nason:2004rx,Frixione:2007vw,Alioli:2010xd,Hartanto:2015uka} generator at NLO with the \NNPDF[3.0nlo] PDF set. The events were interfaced to \PYTHIA[8.2] using the A14 tune and the \NNPDF[2.3lo] PDF set.

The rest of the SM backgrounds were simulated with the \SHERPA[2.2]~\cite{Bothmann:2019yzt} generator. Matrix elements at LO or NLO accuracy in QCD were calculated with the \comix~\cite{Gleisberg:2008fv} and \openloops~\cite{Buccioni:2019sur,Cascioli:2011va,Denner:2016kdg} libraries. The matrix element calculations were matched and merged with the \sherpa\ parton shower based on Catani--Seymour dipole factorisation~\cite{Gleisberg:2008fv,Schumann:2007mg} using the MEPS@NLO prescription~\cite{Hoeche:2011fd,Hoeche:2012yf,Catani:2001cc,Hoeche:2009rj}.
These background samples were simulated using the \NNPDF[3.0nnlo] PDF set, along with the dedicated set of tuned parton-shower parameters developed for {\sherpa}.

For the production of $V\gamma$ ($W\gamma$ and $Z\gamma$) and prompt single-photon ({\gjet}) final states, \SHERPA[2.2.2] was used (without NLO electroweak corrections). Matrix elements at NLO for up to one (two) additional parton emission(s) and at LO accuracy for up to three (four) additional parton emissions were calculated for the $V\gamma$ ({\gjet}) samples.
Diphoton and $V\gamma\gamma$ ($W\gamma\gamma$ and $Z\gamma\gamma$) events were simulated with \SHERPA[2.2.4]. In this set-up, matrix elements at LO accuracy in QCD for up to three additional parton emissions were matched and merged with the \sherpa\ parton shower. To avoid any double counting of events when considering both the $V\gamma$ and $V\gamma\gamma$ samples, duplicated processes were removed, focusing on $V\gamma$ events with a photon from QED final-state radiation.

The contribution from other backgrounds including diboson $VV\gamma$ SM processes are found to be negligible.
Backgrounds due to jets or electrons misidentified as photons are estimated via data-driven techniques and are discussed in Section~\ref{sec:background}.

The SM background samples were processed with a full ATLAS detector simulation~\cite{SOFT-2010-01} based on \GEANT~\cite{Geant4}. The SUSY signal samples were passed through \textsc{ATLFAST-II}, a fast simulation of the ATLAS detector response based on a parameterisation of the performance of the ATLAS EM and hadronic calorimeters and on \GEANT elsewhere. MC events were reconstructed with the same algorithms used for data. An event-by-event reweighting was applied to all MC samples to realistically model the LHC conditions during data collection. This reweighting matches the simulated distribution of the number of inelastic $pp$ collisions per bunch crossing (pile-up) to the one observed in data. The effect of pile-up was modelled by overlaying each hard-scattering event with simulated inelastic $pp$ collisions generated by \PYTHIA[8.186]~\cite{Sjostrand:2007gs} using the \NNPDF[2.3lo] set of PDFs and the A3 tune~\cite{ATL-PHYS-PUB-2016-017}. SM background samples, except the ones that use the \SHERPA\ event generator, were processed with \EVTGEN[1.7.0]~\cite{Lange:2001uf} for the $b$- and $c$-hadron decays. The simulations were further corrected with efficiency scale factors and a smearing of the energy scale of photons, leptons and jets, to better describe the data. Table~\ref{tab:mc_samples_summary} presents a summary of the signal and background samples used in the analysis.

\begin{table}[h!]
\centering
\caption{Details of the SUSY signal and SM background MC samples considered in the analysis.}
\label{tab:mc_samples_summary}
\scriptsize
\begin{tabular}{l|lllrrr}
\toprule
\toprule
Process                  & Generators                     & PDF sets        & Order \\
\midrule
Signal                   & \MADGRAPH[5]\,/\,\PYTHIA[8.2]    & \NNPDF[2.3lo]   & 0,1,2j@LO \\ %
\midrule{\ttgam}                 & \MGNLO\,/\,\PYTHIA[8.2]          & \NNPDF[3.0nlo]\,/\,\NNPDF[2.3lo]    & NLO  \\ %
{\tthgamgam}             & \powheg\,/\,\Pythia[8.2]         & \NNPDF[3.0nlo]\,/\,\NNPDF[2.3lo]   & NLO  \\ %
{\Vgamma}                & \SHERPA[2.2.2]                 & \NNPDF[3.0nnlo] & 0,1j@NLO + 2,3j@LO \\ %
{\gjet}                  & \SHERPA[2.2.2]                 & \NNPDF[3.0nnlo] & 1,2j@NLO + 3,4j@LO \\ %
{\diphoton/\vgammagamma} & \SHERPA[2.2.4]                 & \NNPDF[3.0nnlo] & 0,1,2,3j@LO \\ %
\bottomrule
\bottomrule
\end{tabular}
\end{table}


%
%
\section{Event reconstruction}
\label{sec:objects}
%

Primary vertices are formed from sets of two or more charged-particle tracks, each with transverse momentum $\pt > 0.5 ~\gev$ that are consistent with having originated at the same three-dimensional space point within the luminous region of the colliding proton beam~\cite{ATL-PHYS-PUB-2015-026}. When more than one such primary vertex is found the vertex with the largest scalar sum of the squared transverse momenta of the associated tracks is chosen.

The offline electron and photon reconstruction uses dynamic, variable-size clusters of energy deposits
measured in topologically connected EM and hadronic calorimeter cells~\cite{PERF-2014-07}, called `topo-clusters'. These are matched to reconstructed charged-particles tracks from the ID. Topo-clusters that are not matched to a track or a conversion vertex are classified as unconverted photons. On the other hand, if
a topo-cluster is matched to a conversion vertex, it is classified as a converted photon. Finally, when a
topo-cluster is matched to a track, it is classified as an electron~\cite{EGAM-2018-01}.

Photon candidates are required to satisfy the `tight' identification criteria for the lateral and longitudinal shower
shape~\cite{EGAM-2018-01}, have $\pt > 50$~\gev\ and $|\eta| < 2.37$, and are removed if they are within the
EM calorimeter (ECAL) barrel--endcap transition region defined by $1.37<|\eta|<1.52$.
To reduce the background from jets that could be misidentified as photons, both track and calorimetric isolation requirements
are applied to signal-region photon candidates. The calorimetric isolation energy, {\etiso}, is computed as
the sum of the topo-clusters' transverse energies~\cite{PERF-2014-07} calibrated at the EM scale
within a cone of size $\Delta R = 0.4$ around the cluster barycentre. This {\etiso} is required to be less than
$2.45~\gev + 0.022 \pt$, where \pt is that of the photon. The track isolation variable, \ptIsol, is obtained as the scalar sum
of the transverse momenta of good-quality tracks inside a cone of size $\Delta R = 0.2$ around the candidate, and is required to be less than 0.05$\pt$.

Although electron and muon candidates are not used as part of the signal-region event selection they are included in the definition of some of the control and validation region selections.

Electron candidates are required to have $\pt>10~\gev$ and $|\eta|<2.47$, and to originate from the primary vertex in both the $r$--$z$ and $r$--$\phi$ planes. Longitudinal and transverse impact parameters ($z_{0}$ and $d_{0}$) relative to the primary vertex are required to satisfy $|z_{0} \sin \theta| < 0.5$ mm and $|d_{0}|/\sigma_{d_{0}} < 5$, where $\sigma_{d_{0}}$ is the uncertainty in $d_{0}$.
A `tight' set of identification criteria is imposed~\cite{EGAM-2018-01} that is based on the characteristics of the EM shower development, the quality of the associated reconstructed track, and the angular proximity of the track to the calorimeter energy deposition.
Electrons are required to satisfy a `loose' isolation criterion~\cite{EGAM-2018-01}, and not be within the ECAL barrel--endcap transition region.

Muons are reconstructed by combining compatible track information from the MS and the ID\@. Muon candidates are required to have $\pt>10~\gev$ and $|\eta|<2.7$, to satisfy the `medium' quality criteria~\cite{MUON-2018-03}, to originate from the primary vertex in both the $r$--$z$ and $r$--$\phi$ planes (with impact parameters satisfying $|z_{0} \sin \theta| < 0.5$ mm and $|d_{0}|/\sigma_{d_{0}} < 3$) and to satisfy a `loose' isolation requirement~\cite{MUON-2018-03}.

Jets are reconstructed using the anti-$k_t$ algorithm~\cite{Cacciari:2008gp,Fastjet} with a radius parameter $R = 0.4$, using the information collected in both calorimeters and track detectors~\cite{PERF-2015-09}. The expected average energy contribution from pile-up interactions is subtracted according to the jet area.

Track-based selection requirements are applied to reject jets with $\pt< 60~\gev$ and $|\eta|< 2.4$ that originate from pile-up interactions~\cite{PERF-2014-03}. Jets are kept if they are in the central, $|\eta| < 2.8$, region of the detector and have $\pt>20~\gev$. For the \MET\ computation only (defined in the following) a broader selection including up to $|\eta|<4.5$ is considered. Although jets containing $b$-hadrons ($b$-jets) are not explicitly used in the signal-region selections, they are used in the definition of control regions from which the {\wgamma} and \ttbargam\ MC normalisation factors are extracted, as described in \Sec~\ref{sec:selection}. The $b$-jets are selected from the selected jets within the ID acceptance using the same $\pt$ requirement, and identified by the DL1r algorithm~\cite{FTAG-2018-01}. This algorithm uses the long lifetime, high decay multiplicity, hard fragmentation and large mass of $b$-hadrons to distinguish them from light-flavour jets (jets originated from light quarks and gluons). The $b$-tagging algorithm working point is chosen so that it has a nominal efficiency of 77\% for $b$-jets in simulated \ttbar events and a corresponding probability of misidentifying light-flavour jets below 1\%~\cite{FTAG-2020-08,FTAG-2018-01,FTAG-2019-02}.

The missing transverse momentum is computed with an object-based algorithm considering objects satisfying baseline requirements~\cite{SUSY-2018-11}. Calorimeter energy deposits are matched to high-\pt\ objects in the following order: electrons, photons, jets and muons. Primary-vertex tracks not associated with any such objects are included in the `soft term'~\cite{JETM-2020-03} contribution to \met. The \ETM\ is computed from the negative vector sum of the transverse momenta of calibrated reconstructed physical objects and the soft term.

Due to possible final-state object misidentification, a single object can be reconstructed in more than one way and thus effectively counted multiple times. A procedure to remove these overlaps is applied to preselected objects before the corresponding isolation requirements are imposed. The basic strategy and the order of removal is described in Refs.~\cite{SUSY-2016-04,ATLAS:2020uiq}.


%
\section{Event selection}
\label{sec:selection}
%

This analysis is performed using the full Run-2 data sample of LHC proton--proton collisions at \ECMS\ collected by the ATLAS detector between 2015 and 2018, corresponding to an integrated luminosity of \lumi\ after the application of beam, detector and data quality requirements~\cite{DAPR-2018-01}.

Events were selected using the single-photon trigger ~\cite{TRIG-2018-05}, and are required to have at least one photon candidate reconstructed offline with $\pt>145~\gev$. The trigger was fully efficient for the signal selection requirements, as described in~\cite{EGAM-2018-01}.

A set of signal regions sensitive to the targeted SUSY signatures are defined for the analysis, discriminating against SM processes contributing to the same final state. Given the relatively high mass of the electroweakinos targeted by this search the \pt of the final-state photon is expected to be large.

Three discovery signal regions (SRL, SRM and SRH) are defined to target the phase space of the model with low, medium and high {\ninoone} mass. At higher masses, a more energetic gravitino is produced and consequently \MET\ is expected to be larger. The main distinction between the signal regions is therefore the requirement on this variable.

The contribution of \MET  relative to the total energy contribution of all objects present in the event is quantified using the ratio of \MET and the effective mass, ${\meff} = {\met + \ptph + \sum_{i}^{\text{jets}} p_{\text{T}}^{i}}$, where the leading photon is the one with the highest $\pt$. The $\frac{\met}{\meff}$ ratio is expected to be larger for the SUSY signal than for the backgrounds, hence the requirement on this variable in the signal region.

In events characterised by a large reconstructed \MET\ without a significant contribution from non-interacting particles but arising from instrumental sources and poorly reconstructed objects, the \MET\ vector tends to be aligned with either the photon or one of the two leading jets.
A selection based on the transverse angular separations between these objects, $\Delta\phi(\gam, \met)$ between the leading photon and the \MET\ vector and $\Delta\phi(\text{jet}, \met)$ the minimum separation between the leading or sub-leading jet and the \MET\ vector, allows the contributions from these background processes to be reduced. The significance of \MET~\cite{JETM-2020-03} is another discriminant between both sources of \MET, thus a requirement is applied on it to further increase sensitivity. Additional jets are expected to be produced frequently in the signal decay chain, thus at least one jet is required in each signal region. In order to be orthogonal to the analysis focusing on similar final states but with the lightest Higgs boson decaying into photons~\cite{SUSY-2020-17}, events with two photons with their invariant mass within the Higgs boson mass window ($120 < m_{\gamma\gamma} < 130\, \gev$) are excluded. This selection has a negligible impact on the overall sensitivity of the search.

\begin{figure}[ht!]
\begin{center}
\includegraphics[width=0.7\textwidth]{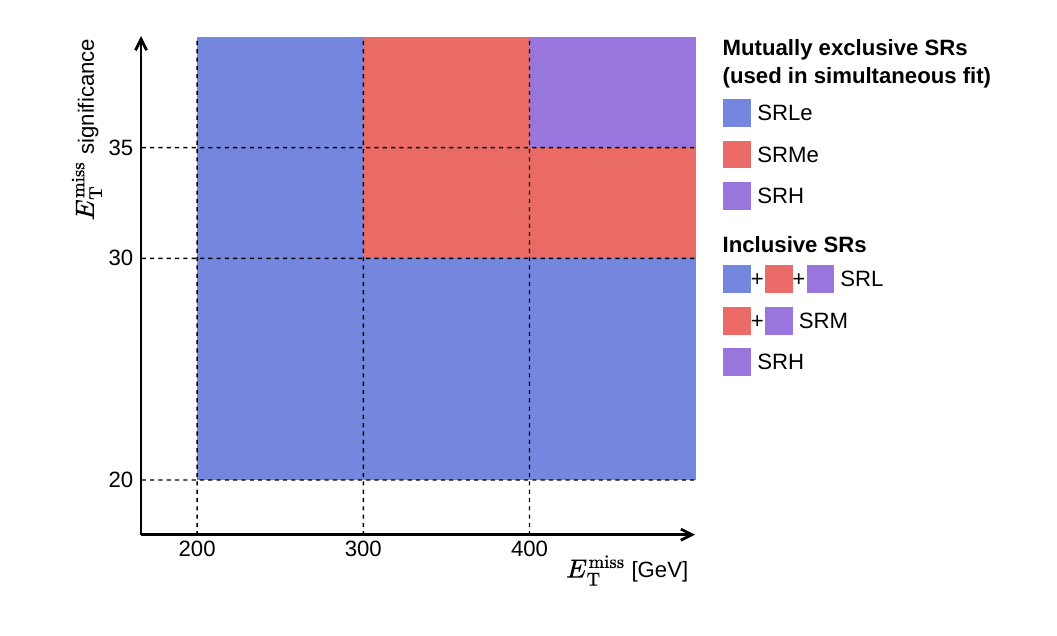}
\caption{Schematic view of the three signal regions, defined in the {\met} and {\met} significance plane, used in the simultaneous fit for exclusion limits (SRLe, SRMe and SRH). The other inclusive SRs can be defined from these as $\text{SRL} = \text{SRLe}\, \cup\, \text{SRMe}\, \cup\, \text{SRH}$ and $\text{SRM} = \text{SRMe}\, \cup\, \text{SRH}$. The dotted lines represent the requirements on the corresponding variables.}
\label{fig:diagram_exclusion_srs}
\end{center}
\end{figure}

The event selection criteria for the full set of signal regions are summarised in Table \ref{tab:sr_selection}, including both the inclusive selections for discovery (SRL, SRM, and SRH) and the orthogonal selections for exclusion (SRLe and SRMe). The additional signal subregions SRLe and SRMe are designed to improve the exclusion limits in the absence of an excess of events in the inclusive signal regions. Compared to the discovery signal regions, the requirements on \MET and \MET significance are tightened in order to make the three signal regions orthogonal (SRLe, SRMe and SRH), allowing a simultaneous exclusion fit. The signal region SRH is used both for discovery and limit setting without any modifications as shown schematically in Figure \ref{fig:diagram_exclusion_srs}.

\begin{table}[ht!]
\centering
\caption{Selections for the discovery signal regions SRL, SRM, SRH, and the exclusion signal regions SRLe and SRMe.}
\begin{tabular}{l|ccc|cc}
\toprule
\toprule
Signal regions & SRL & SRM & SRH & SRLe & SRMe \\
\midrule
\nph & \multicolumn{5}{c}{$>0$} \\
\ptph[\GeV] & \multicolumn{5}{c}{$>145$} \\
\njet & \multicolumn{5}{c}{$>0$} \\
\dphijetmet & \multicolumn{5}{c}{$>0.4$} \\
\dphigammet & \multicolumn{5}{c}{$>0.4$} \\
\met/\meff & \multicolumn{5}{c}{$>0.5$} \\
\myy[\GeV] & \multicolumn{5}{c}{$\notin [120,130]$} \\
\met[\GeV] & $>200$ & $>300$ & $>400$ & $>200\ (\notin \text{SRM})$ & $>300\ (\notin \text{SRH})$ \\
\met significance & $>20$ & $>30$ & $>35$ & $>20\ (\notin \text{SRM})$ & $>30\ (\notin \text{SRH})$ \\
\bottomrule
\bottomrule
\end{tabular}
\label{tab:sr_selection}
\end{table}

%
\section{Background estimation}
\label{sec:background}
%

The background for this search is composed of all SM processes that produce photons, jets and \met in the final state. Both the production of neutrinos or a misreconstruction of the energy of one or more objects can generate \met in the final state. Processes with jets or electrons can emulate the targeted topologies if any of these is misreconstructed as a photon.

The dominant SM background contribution to the SRs is expected to be from \znunugamma, with a lesser contribution from \wgamma and \ttgamma in most regions of the analysis. The simulated predictions for these contributions are normalised using dedicated control regions where the simulations are constrained by the number of data events observed. Other minor processes, such as \tthgamgam, \gjet, \diphoton and \vgammagamma, are estimated directly using the MC simulation. An estimate of the background contribution due to misidentified jets or electrons is determined via data-driven techniques.

Three control regions labelled CRZ, CRW, and CRT are used to obtain the MC normalisation for the \znunugamma, \wgamma and \ttgamma events, respectively. These CRs are designed to be orthogonal but still kinematically similar to SRs, and enhanced in the background process of interest, with negligible signal contamination (less than 5\%).
Since \znunugamma has the same final state as the SUSY signal, the design of a CR is not possible without having a high signal contamination. Therefore, a single CRZ with two electrons or two muons, focusing on the leptonic decays of the $Z$ boson, \zeegamma and \zmumugamma, is used to normalise the three corresponding backgrounds simultaneously in each of the CRs.
Since these processes have no \met in the final state, an alternative reconstruction of \met, removing the leptons present in the event, is used to construct the CRs kinematically closer to the SRs.
The CRT region is defined by requiring a photon, a lepton, jets and at least two $b$-tagged jets, with a looser {\met} requirement to increase the yields.
The CRW region is defined by requiring a photon, a lepton and a $b$-jet veto to mitigate the contamination from \ttbargam, and a requirement on the transverse mass of the lepton and \met, \mtlep \footnote{$m_{\text{T}}=\sqrt{2p_{\text{T}}^{l}E_{\text{T}}^{\text{miss}}(1-\cos[\Delta \phi (p_{\text{T}}^{l},p_{\text{T}}^{\text{miss}})])}$}, to increase the background purity.
All CRs invert the requirement on \met/\meff to ensure orthogonality with the SRs.

A further set of event selections define validation regions (VRs) used to verify the results of the background estimation procedure. They are designed to lie kinematically between the signal regions and the control regions, with one or more criteria
inverted or modified to reduce possible signal contamination. In general the requirement on \met/\meff is in accordance with the SRs, and the \met requirement increased to be closer to them.
The VRZ, VRW and VRT are designed to validate the correct estimate of the \znunugamma, \wgamma and \ttgamma backgrounds.

Jets can be misidentified as photons (called `fake photons') if they contain a neutral hadron that carries most of the jet energy and that decays into a pair of collimated photons, resulting in an EM object resembling a single, highly energetic photon.
This background arises primarily from QCD multijets, \wjets\ and semileptonic \ttbar\ events, and it is reduced by applying the `tight' identification criteria to photon candidates. After this requirement, the data sample is expected to be dominated by real photons with moderate jet contamination. This misidentification rate is not modelled accurately in MC simulation, thus a data-driven sideband counting method~\cite{ATLAS:2020uiq} is used. The method makes use of the different isolation profiles expected for real photons and misidentified jets~\cite{STDM-2010-08}. Both tracking and calorimetric isolation variables of the photon candidate, as defined in Section~\ref{sec:objects}, are considered simultaneously. The photon trigger used to collect the data required a looser photon identification than the `tight' offline version. Some photon candidates from misidentified jets fail to satisfy the offline requirement but satisfy an intermediate selection.
These jets, called `non-tight photons', are defined as those candidates satisfying the `loose' identification and the `tight' selection requirements, except for at least one of four criteria associated with energy deposits in the EM calorimeter~\cite{ATLAS:2020uiq}, chosen to be largely uncorrelated with the isolation variables. Hence the use of non-tight photons enhances the contribution of jets misidentified as photons (called `jets faking photons'), as required for this method. %
In the identification--isolation plane, the method defines a signal region A consisting of isolated photon candidates that satisfy the `tight' identification, and three control regions, namely B, C and D, with photon candidates being non-isolated and `tight', isolated and non-tight, and non-isolated and non-tight, respectively. A small contribution of real photons still contaminates the three defined control regions; these are additionally referred to as `leaked-photons'.

To accurately count the number of real, leaked and fake photons in the different regions, a series of template fits to the isolation profiles in data are performed using templates obtained from MC~\cite{EGAM-2018-01}. These templates are fitted using double sided asymmetric Crystal Ball (DSACB) functions (a Gaussian core with asymmetric non-Gaussian power-law tails) ~\cite{Gaiser:1982yw} to the tight and non-tight photon distributions in data. DSACB functions are also used to model the contribution of jets faking photons. The procedure to count the events is as follows.
Firstly, using the leaked-photons template obtained from MC, the fake photons template is obtained by doing a composite fit to the non-tight photons isolation distribution in data. In this way, both the leaked and fake photons components are obtained in the non-tight regions C and D. Secondly, using the fake photon template and the real photon template obtained from MC, a fit to the tight data profile distribution is performed, resulting in the real and fake photons components in regions A and B.
Finally, real, leaked and fake photons components can be extracted from the distributions by integrating the shapes in the isolated and non-isolated ranges.

For this method to be applicable, the correlation between identification and isolation for jets misidentified as photons must be small. Deviations from this expected behaviour are corrected by computing a factor that takes the residual correlation into account~\cite{ATLAS:2020uiq}. It is computed in auxiliary ABCD regions (defined in a similar way as before but where the calorimetric isolation requirement fails) as the ratio \(N_\text{A} N_\text{D} / N_\text{B} N_\text{C}\), where \(N_\text{X}\) is the number of jets faking photons in the auxiliary ABCD regions. The residual correlation is found to be less than 20\%. Furthermore, corrections to the method using the number of leaked-photons are applied. The systematic uncertainties in the method are evaluated by varying the definition of the non-tight objects, and considering the differences introduced by the residual correlation between the regions.

Electrons and positrons can also be misidentified as photons, and when they have high-\pt\ there can be significant contamination from SM processes such as $W/Z$\,+\,jets and \ttbar\ in the signal regions.
This background is estimated by weighting the number of electron events observed in electron control samples by the electron-to-photon fake rate.
Electron control samples are taken from the same control, validation and signal regions as used in the analysis, but the photon kinematic selections are applied to electrons, and then a high-$\pt$ isolated electron is required and signal photons are vetoed.
To estimate the electron-to-photon fake rate, a method based on a sample of \zee\ data events is used~\cite{SUSY-2016-04,ATLAS:2020uiq}. Since the $Z$ boson cannot decay directly into an electron and a photon, the electron--photon events appearing under the $Z$ peak most likely correspond to misidentified electrons. However, the same applies to other particles decaying into pairs of electrons that could contribute to the background around the $Z$ mass values.
Therefore, these background contributions are subtracted by also taking into account the contamination from random combinatorics background. Also, only events with $\etmiss<40~\gev$ are selected, to avoid electrons from the $W$ decays.
The electron-to-photon fake factor is then estimated as the ratio of the number of electron--photon pairs to the number of electron--electron pairs found under the $Z$ peak when fitting the invariant mass distribution. This fit uses a double sided Crystal Ball (DSCB) function to model the $Z$ peak, and a Gaussian distribution to model the small non-resonant backgrounds to \zee\ production. Only the pairs within a defined invariant mass window are selected to compute the electron-to-photon fake factor. This window is defined as $\pm3\sigma$ around the centre of the peak in the DSCB function, where $\sigma$ is the width of the peak. Systematic uncertainties are evaluated by varying the Z-mass integration range, the energy calibration, and recalculating the fake factor without background subtraction. The background estimation is then performed by applying the electron-to-photon misidentification rate to an electron-control sample. This control sample mirrors the original control, validation, and signal regions of the analysis but replaces photon selections with analogous electron criteria. Specifically, events are required to contain a high‑\pt, isolated electron (while vetoing genuine photons), and all other kinematic selections used for photons are applied to electrons. The prediction of the background due to electrons misidentified as photons is validated in a dedicated validation region (VRE) based on the calculated fake factors. It mostly targets $W$($e\nu$)\,+\,jets events (by requiring low \mtgam), where a boosted $W$ boson (including those coming from top quarks) decays into a neutrino (giving high {\MET}) and an almost collinear high-{\pt} electron (misidentified as a photon).

The selection criteria for the CRs and VRs are presented in Table \ref{tab:CR_VR_selection}.

\begin{table}[ht!]
\centering
\caption{Selection criteria for the control regions and the validation regions used for the $W$, top, $Z$, and fakes background estimation. All CRs and VRs also include the following common cuts: $\nph>0$, $\ptph>145 ~ \gev$ and $\njet>0$.
The asterisk $^{*}$ means that \met in this case is reconstructed without considering the leptons. Additionally in  regions CRZ and VRZ an $\met <50$ selection is applied in the variable including leptons.
}
\resizebox{\textwidth}{!}{
\begin{tabular}{l|ccc|cccc}
\toprule
\toprule
Variable                & CRW    & CRT      & CRZ         & VRW    & VRT      & VRZ         & VRE    \\
\midrule
\nlep                  & $1$    & $\ge1$   & $2$         & $1$    & $\ge1$   & $2$         & $-$    \\
\nbjet                 & $0$    & $\ge2$   & $-$         & $-$    & $\ge1$   & $-$         & $-$    \\
\dphijetmet            & $>0.4$ & $>0.4$   & $>0.4^{*}$  & $>0.4$ & $>0.4$   & $>0.4^{*}$  & $>0.4$ \\
\dphigammet            & $>0.4$ & $>0.4$   & $>0.4^{*}$  & $>0.4$ & $>0.4$   & $>0.4^{*}$  & $<0.4$ \\
\met[GeV]             & $>200$ & $>150$   & $>100^{*}$  & $>200$ & $>200$   & $>200^{*}$  & $>200$ \\
\met/\meff             & $<0.35$& $<0.35$  & $<0.35^{*}$ & $>0.35$& $>0.35$  & $>0.35^{*}$ & $>0.25$\\
\met significance      & $-$    & $-$      & $-$         & $<20$  & $<20$    & $-$         & $>10$  \\
\mtlep[GeV]           & $<100$ & $-$      & $-$         & $-$    & $-$      & $-$         & $-$    \\
\mtgam[GeV]           & $-$    & $-$      & $-$         & $-$    & $-$      & $-$         & $<80$  \\
\bottomrule
\bottomrule
\end{tabular}
}
\label{tab:CR_VR_selection}
\end{table}

Likelihood fits \cite{Baak:2014wma} are performed assuming (i) a background-only hypothesis on all CRs to estimate the total background in the SRs and VRs; (ii) a model-dependent signal plus background hypothesis where the fit is performed in the CRs and independent SRs simultaneously; and (iii) a model-independent signal plus background hypothesis, where both the CRs and individual SRs are used in the same manner as for the model-dependent signal fit with the number of signal events in the SRs added as a parameter to the fit. This approach constrains the expected background to the yields observed in the data using the CRs and reduces the systematic uncertainties.%

\begin{figure}[ht!]
\centering
\includegraphics[width=0.85\textwidth]{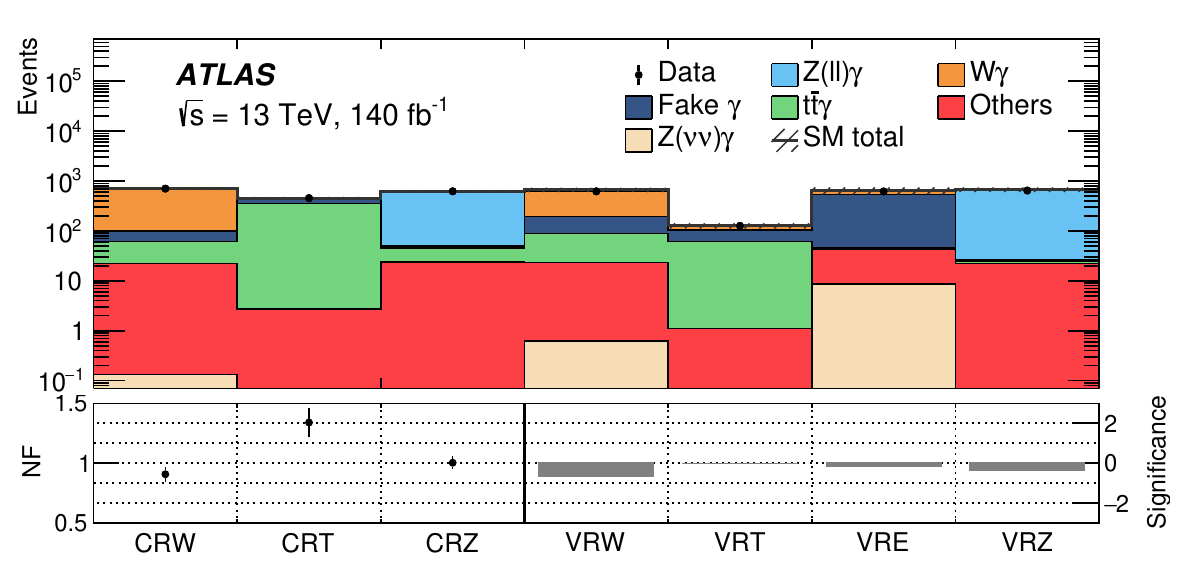}
\caption{The observed and expected yields after the background-only fit in the control and validation regions. The lower panel shows the background normalisation factors (NF) estimated in each control region, and the significance, in standard deviations, between the observed and expected yields in the validation regions, considering both the systematic and statistical uncertainties in the background expectation. The significance shown is calculated with the profile likelihood method from~\cite{Cowan:2010js}, adding a minus sign if the yield is below the prediction. Minor background processes, such as \gjet, \diphoton and \vgammagamma, are included under `Others'.}
\label{fig:regions_pulls}
\end{figure}

Figure \ref{fig:regions_pulls} shows the data event yields and the SM expectation in the different control and validation regions obtained with a background-only maximum-likelihood fit, constraining the normalisation of the dominant backgrounds and including those estimated by using data-driven techniques. The lower panel shows the statistical significance, in standard deviations, between the observed and expected yields in the VRs~\cite{Choudalakis2012} and the normalisation factors for each CRs. The normalisation factors are close to unity, except for the CRT region (1.34 $\pm$ 0.12), in agreement with previous results~\cite{SUSY-2018-11}. Good agreement is found between data and SM background predictions in all validation regions.

%
\section{Systematic uncertainties}
%
%

All background processes estimated either by making use of MC simulations or by data-driven methods, as well as MC signal predictions, are affected by systematic uncertainties that mainly originate from two kinds of sources:
experimental and theoretical ones. These uncertainties can impact the expected yields in the analysis regions and are introduced in the fits as nuisance parameters.

The uncertainty in the combined 2015--2018 integrated luminosity is 0.83\%~\cite{DAPR-2021-01}, obtained using the LUCID-2 detector~\cite{Avoni_2018} for the primary luminosity measurements.
The uncertainty in the pile-up reweighting is also considered.

The systematic uncertainties due to the photon identification and isolation efficiencies are estimated following the prescriptions in Ref.~\cite{EGAM-2018-01}. They are evaluated by varying the correction factors for the photon selection efficiencies in MC simulation by the corresponding uncertainties. The photon energy scale is determined using samples of $Z\to ee$ events, varying the scale corrections and resolutions upwards and downwards by one standard deviation~\cite{EGAM-2018-01}.
For electrons~\cite{EGAM-2018-01} and muons~\cite{PERF-2015-10}, similarly to photons, the uncertainties from the identification efficiencies, energy scales and resolutions are determined from $Z\to ll$ and $\jpsi \to ll$ control samples, where $l$ denotes electrons or muons.

For jets, the energy scale and resolution uncertainties are derived following the procedure described in Ref.~\cite{PERF-2016-04}, using a simplified scheme with 38 parameters.
A set of $b$-tagging uncertainties is also considered, taking an envelope around the nominal jet weight for the selection of different flavour jets~\cite{FTAG-2020-08,FTAG-2018-01,FTAG-2019-02}.

For {\MET}, the uncertainties of the constituent objects are propagated through the calculation, and additional uncertainties accounting for the scale and resolution of the soft term~\cite{JETM-2020-03} are considered.

There are two different kinds of uncertainties affecting the estimate of the background due to jets/$e^{\pm}$ misidentified as photons: the systematic uncertainty from the method used to estimate the fake factors and the
statistical uncertainty in the control sample (see Section~\ref{sec:background}).

For each of the main simulation background samples ($Z(\nu\nu)\gamma$, $W\gamma$, $Z(ll)\gamma$ and $t\bar{t}\gamma$) and signals, different sources of theory uncertainties are assessed.
For renormalisation and factorisation scales $\muR$ and $\muF$ three independent nuisance parameters are used: two varying each of the scales independently by a factor of two, and one as a coherent variation of both scales.
The PDF uncertainty is taken from an envelope of the nominal PDF (\NNPDF[3.0]) and its variations~\cite{Butterworth:2015oua}. Finally, uncertainties associated with the $\alphas$ determination are considered.

The relative impact of each systematic uncertainty in the SM background expectation after the background-only fit is presented in Table~\ref{tab:syst_rel_impact}.

\begin{table}[ht!]
\centering
\caption{Summary of the different systematic uncertainties in the SM background expectation (in \%) for the different SRs after the background-only fit. The individual uncertainties can be correlated and do not necessarily add in quadrature to equal the total (stat.\ + syst.) uncertainty.}
\resizebox{\textwidth}{!}{
\begin{tabular}{lccccc}
\toprule
& SRL [\%]    & SRM [\%]  & SRH [\%] & SRLe [\%] & SRMe [\%] \\
\midrule
Total (stat.\ + syst.) uncertainty                    & 8.8         & 10.0~~    & 13.3~~   & 8.9       & 10.9~~    \\
MC statistical uncertainty                            & 1.7         & 5.0       & 7.5      & 1.9       & 6.9       \\
\midrule
MC theory                                             & 6.0         & 6.0       & 5.9      & 6.0       & 6.1       \\
Jet energy resolution, scale, and pile-up rejection   & 4.1         & 3.8       & 7.7      & 4.2       & 1.9       \\
Photon identification and isolation                   & 1.7         & 1.3       & 1.7      & 1.8       & 1.1       \\
Pile-up reweighting                                   & 1.7         & < 1~~~~~~ & 1.3      & 1.8       & < 1~~~~~~ \\
Muon reconstruction and identification                & 1.6         & 1.3       & 1.6      & 1.8       & < 1~~~~~~ \\
Jet $b$-tagging calibration                           & 1.6         & 1.2       & 1.6      & 1.8       & < 1~~~~~~ \\
Fake photons from jets                                & 1.6         & < 1~~~~~~ & 1.7      & 1.7       & 1.6       \\
Electron reconstruction, identification, and isolation& 1.6         & 1.2       & 1.6      & 1.7       & 1.0       \\
Fake photons from electrons                           & 1.6         & 1.7       & 2.1      & 1.5       & 1.6       \\
$E_{\mathrm{T}}^{\text{miss}}$ soft-term resolution and scale  & 1.3         & 1.3       & 1.7      & 1.3       & 1.1       \\
Electron/photon energy resolution and scale           & < 1~~~~~~   & 2.1       & < 1~~~~~~& 1.3       & 3.1       \\
Luminosity                                            & < 1~~~~~~   & < 1~~~~~~ & < 1~~~~~~& < 1~~~~~~ & < 1~~~~~~ \\
\bottomrule
\end{tabular}
}
\label{tab:syst_rel_impact}
\end{table}


%
\section{Results}
%
%

A background-only fit is performed, as discussed in Section~\ref{sec:background}, to evaluate the SM background expectations in all regions of the analysis. The observed event yield in each CR is used as a constraint in the fit to adjust the corresponding background normalisation, assuming that no signal contributes to this yield. The systematic uncertainties in the expected values are included in the fit as nuisance parameters, constrained by Gaussian or log-normal distributions with widths corresponding to the sizes of the uncertainties. They are treated as correlated, when appropriate, between the various regions. The results of the fit are validated in the VRs as shown previously in Figure  \ref{fig:regions_pulls}.

\begin{figure}[ht!]
\centering
\includegraphics[width=0.85\textwidth]{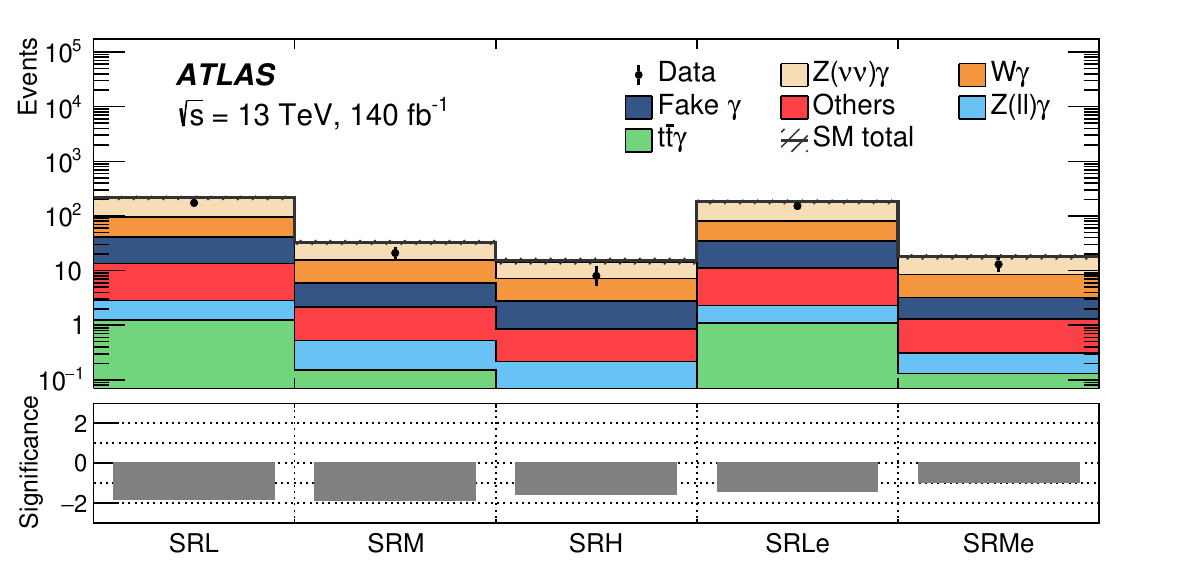}
\caption{The observed and expected yields in all the signal regions. The lower panel shows the difference, in standard deviations, between the observed and expected yields considering both the systematic and statistical uncertainties in the background expectation.}
\label{fig:sr_pulls}
\end{figure}

The observed and expected yields in the SRs are presented in Table \ref{tab:fit_result_sr}, discriminating the different background contributions (minor backgrounds like \gjet, \diphoton and \vgammagamma, are grouped into a single label `Others'). No excess is found above the predicted SM background. The comparisons are also shown in Figure \ref{fig:sr_pulls}.

The predictions slightly exceed the observed data, though the difference is not statistically significant, remaining below two standard deviations. After obtaining these results, further studies were carried out to assess possible mismodelling of the diboson or fake-photon backgrounds, but no evidence of any anomalies was observed. A similar trend was observed in an ATLAS $Z$ cross-section study~\cite{STDM-2017-18} and in the equivalent CMS GGM SUSY search~\cite{CMS-SUS-21-009}.

\begin{figure}[ht!]
\centering
\begin{subfigure}[h]{0.49\linewidth}
\centering
\includegraphics[width=\textwidth]{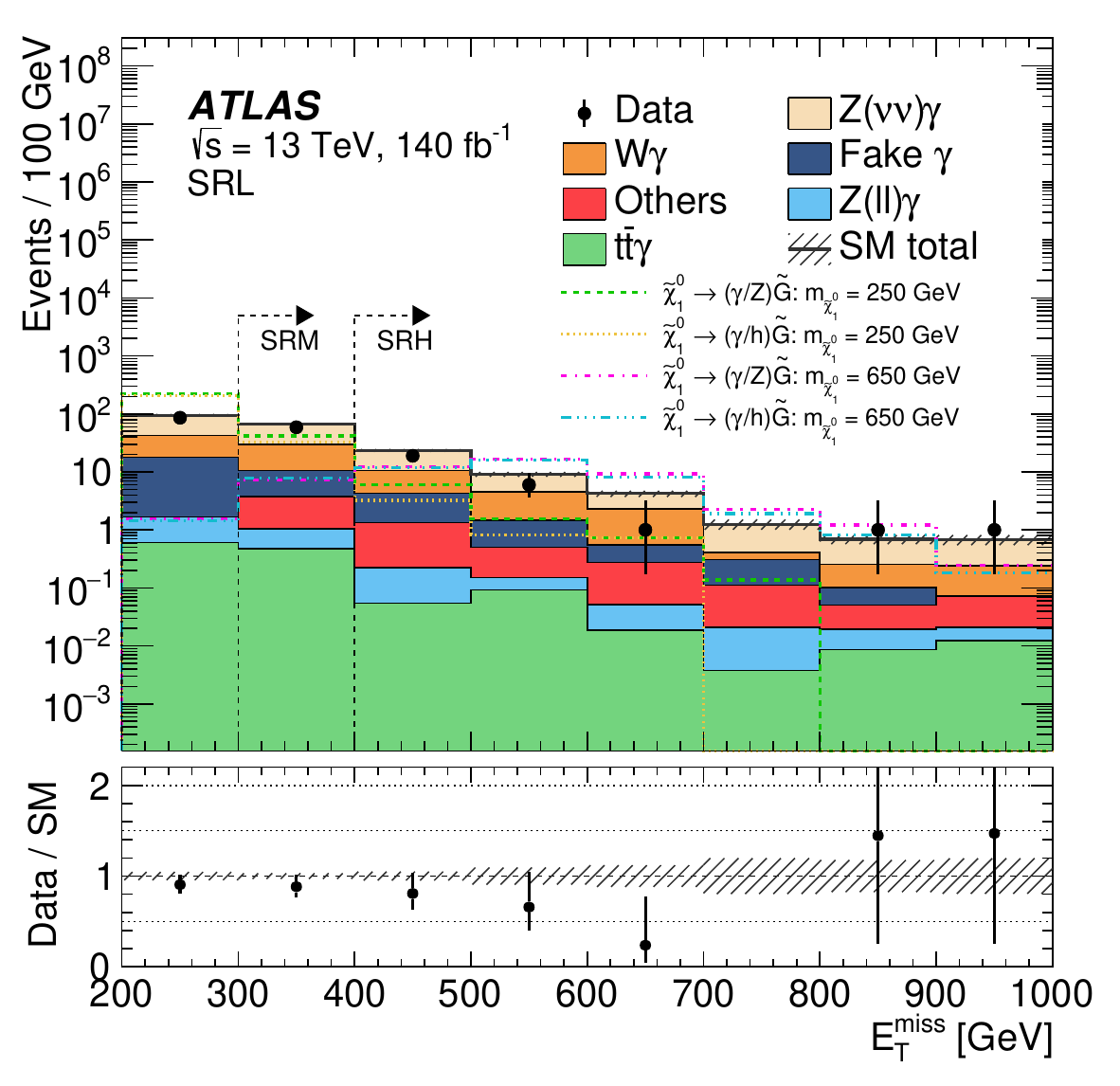}
\caption{}
\end{subfigure}
\begin{subfigure}[h]{0.49\linewidth}
\includegraphics[width=\textwidth]{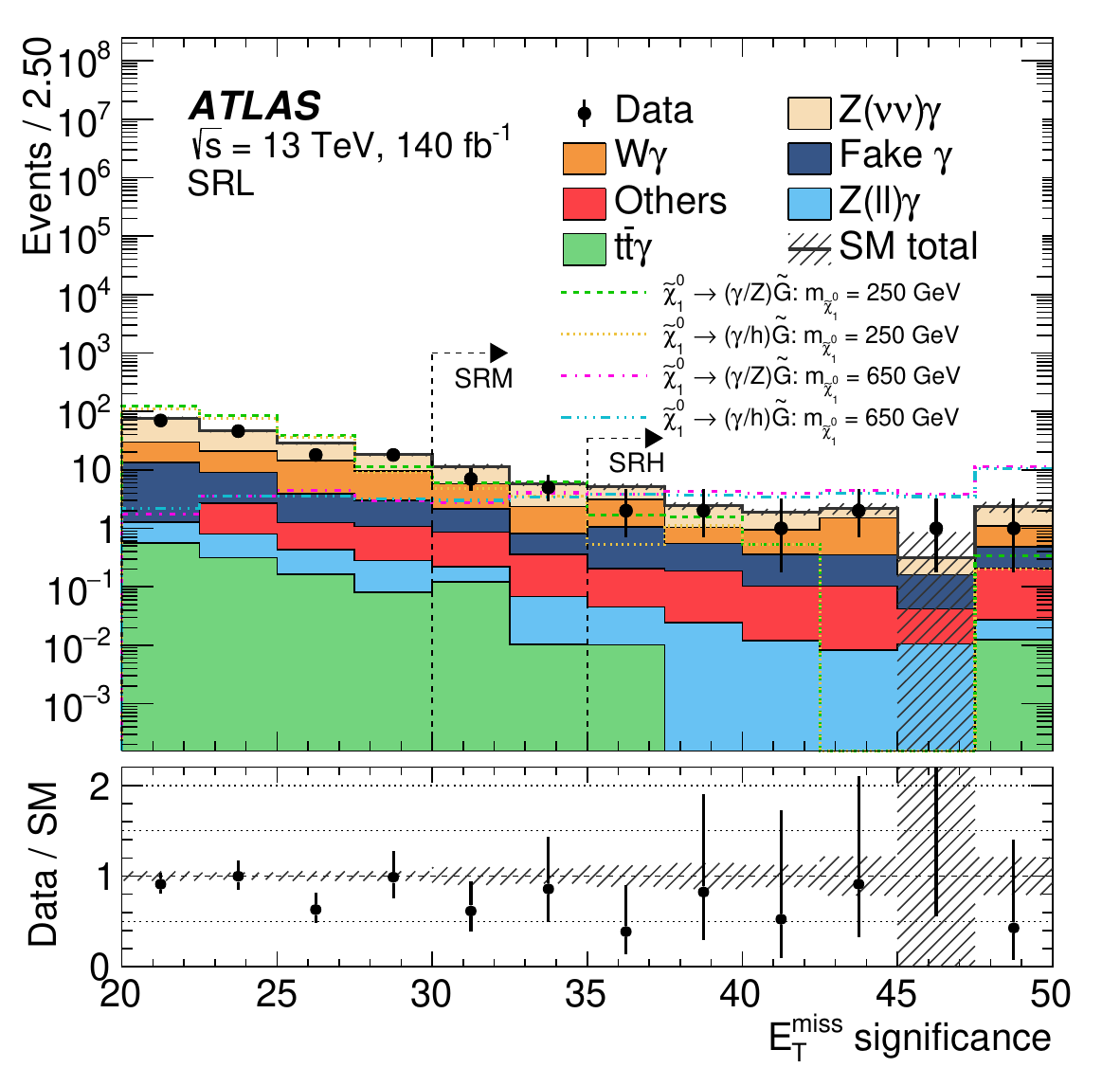}
\caption{}
\end{subfigure}
\caption{Observed distributions in the signal region SRL after the background-only fit for (a) {\met} and (b) {\met} significance variables. The \MET\ and \MET\ significance requirements (that simultaneously define the other signal regions, SRM and SRH), are indicated in the distributions. The uncertainties in the SM background are statistical only. The last bin includes overflow. The figure shows predictions for different models and masses of neutralinos.}
\label{fig:dist_srl_bkgonly_unblinded}
\end{figure}

Figure \ref{fig:dist_srl_bkgonly_unblinded} shows the observed distributions of \MET\ and its significance in the discovery signal region (SRL), compared with the expected background.

\begin{table}
\centering
\small
\caption{Observed events, background estimates after the background-only fit, and model-independent limits in all the signal regions. The background uncertainties are both systematic and statistical. Minor backgrounds like \gjet, \diphoton and \vgammagamma are grouped into `Others'.
The model-independent fit results include the 95\% confidence level (CL) upper limits on the visible cross-section ($\langle\epsilon\sigma\rangle_{\text{obs}}^{95}$), the number of signal events ($S_{\text{obs}}^{95}$), and the number of signal events given the expected number (and $\pm 1\sigma$ excursions on the expectation) of background events ($S_{\text{exp}}^{95}$). The discovery $p$-value is in all cases $p_0 = 0.50$ and its associated significance $Z = 0$.}

\begin{tabular}{lS@{\(\,\pm\,\)}SS@{\(\,\pm\,\)}SS@{\(\,\pm\,\)}SS@{\(\,\pm\,\)}SS@{\(\,\pm\,\)}S}
\toprule
\textbf{Signal region}   &
\multicolumn{2}{c}{SRL}  &
\multicolumn{2}{c}{SRM}  &
\multicolumn{2}{c}{SRH}  &
\multicolumn{2}{c}{SRLe} &
\multicolumn{2}{c}{SRMe} \\
\midrule
Observed events &
\multicolumn{1}{S@{\(\,\phantom{\pm}\,\)}}{173} & &
\multicolumn{1}{S@{\(\,\phantom{\pm}\,\)}}{21}  & &
\multicolumn{1}{S@{\(\,\phantom{\pm}\,\)}}{8}   & &
\multicolumn{1}{S@{\(\,\phantom{\pm}\,\)}}{152} & &
\multicolumn{1}{S@{\(\,\phantom{\pm}\,\)}}{13}  & \\

\midrule
Expected SM events  &
214   & 19   &
32.5  & 3.3  &
14.6  & 2.0  &
182   & 16   &
17.9 & 2.0 \\

\midrule

$Z(\nu\nu)\gamma$  &
116  & 14   &
16.4 & 2.2  &
7.2  & 1.1  &
100  & 12   &
9.2  & 1.4  \\

$W\gamma$  &
57   & 8     &
10.0 & 1.6   &
4.6  & 1.1   &
47   & 7     &
5.5  & 0.9   \\

Fake $\gamma$ &
28   & 5      &
3.9  & 0.9    &
2.0  & 0.5    &
24   & 5      &
1.9  & 0.5    \\

$Z(ll)\gamma$ &
1.60 & 0.21   &
0.38 & 0.10   &
0.20 & 0.07   &
1.22 & 0.23   &
0.18 & 0.10   \\
$t\bar{t}\gamma$                                           &
1.27 & 0.29       &
0.15 & 0.04       &
0.02 & 0.02       &
1.12 & 0.26       &
0.13 & 0.04       \\
Others                                                     &
10.7 & 0.8        &
1.64 & 0.23       &
0.64 & 0.07       &
9.0  & 0.9        &
1.00 & 0.21       \\

\midrule

$\langle A \epsilon \sigma \rangle_{\text{obs}}^{95}$ [fb] &
\multicolumn{1}{S@{\(\,\phantom{\pm}\,\)}}{0.15} & &
\multicolumn{1}{S@{\(\,\phantom{\pm}\,\)}}{0.05} & &
\multicolumn{1}{S@{\(\,\phantom{\pm}\,\)}}{0.04} & &
\multicolumn{1}{S@{\(\,\phantom{\pm}\,\)}}{0.15} & &
\multicolumn{1}{S@{\(\,\phantom{\pm}\,\)}}{0.05} & \\
$S_{\text{obs}}^{95}$                            &
\multicolumn{1}{S@{\(\,\phantom{\pm}\,\)}}{20.6} & &
\multicolumn{1}{S@{\(\,\phantom{\pm}\,\)}}{7.2}  & &
\multicolumn{1}{S@{\(\,\phantom{\pm}\,\)}}{5.4}  & &
\multicolumn{1}{S@{\(\,\phantom{\pm}\,\)}}{21.3} & &
\multicolumn{1}{S@{\(\,\phantom{\pm}\,\)}}{7.0}  & \\
$S_{\text{exp}}^{95}$                            &
\multicolumn{1}{S@{\(\,^{+}_{-}\,\)}}{44}        & \multicolumn{1}{l}{\hspace{-3pt}$^{17}_{12}$}    &
\multicolumn{1}{S@{\(\,^{+}_{-}\,\)}}{15}        & \multicolumn{1}{l}{$^{7}_{4}$}      &
\multicolumn{1}{S@{\(\,^{+}_{-}\,\)}}{10.1}      & \multicolumn{1}{l}{\hspace{-0pt}$^{5.0}_{3.1}$}&
\multicolumn{1}{S@{\(\,^{+}_{-}\,\)}}{40}        & \multicolumn{1}{l}{\hspace{-3pt}$^{16}_{11}$}    &
\multicolumn{1}{S@{\(\,^{+}_{-}\,\)}}{11.1}      & \multicolumn{1}{l}{\hspace{-0pt}$^{5.4}_{3.4}$}\\
\bottomrule
\end{tabular}

\label{tab:fit_result_sr}
\end{table}

For each SR, model-independent upper limits at 95\% confidence level (CL) are set on the number of observed ($S_{\text{obs}}^{95}$) and expected ($S_{\text{exp}}^{95}$) events, from any scenario of physics beyond the SM that could produce an excess in the final state considered. The limits are obtained using the CL$_\text{s}$ prescription in the asymptotic approximation~\cite{Read:2002hq,Cowan:2010js}. In Table~\ref{tab:fit_result_sr} the upper limits on the number of events are presented together with
the 95\% CL upper limit on the visible cross-section $\sigma \times A \times \epsilon$ (denoted $\langle A \epsilon\sigma \rangle_{\text{obs}}^{95}$), obtained by normalising the upper limit on the number of signal events to the integrated luminosity, where $\sigma$ is the production cross-section for a beyond-the-SM signal, $A$ is the acceptance (fraction of events with objects satisfying all the kinematic selections at particle level) and $\epsilon$ is the efficiency (fraction of observed events after detector reconstruction). For all the SRs, the discovery $p$-value ($p_0$) is capped at 0.5 because the predictions exceed the observation.

\begin{figure}
\centering
\begin{subfigure}[h]{0.49\linewidth}
\centering
\includegraphics[width=\textwidth]{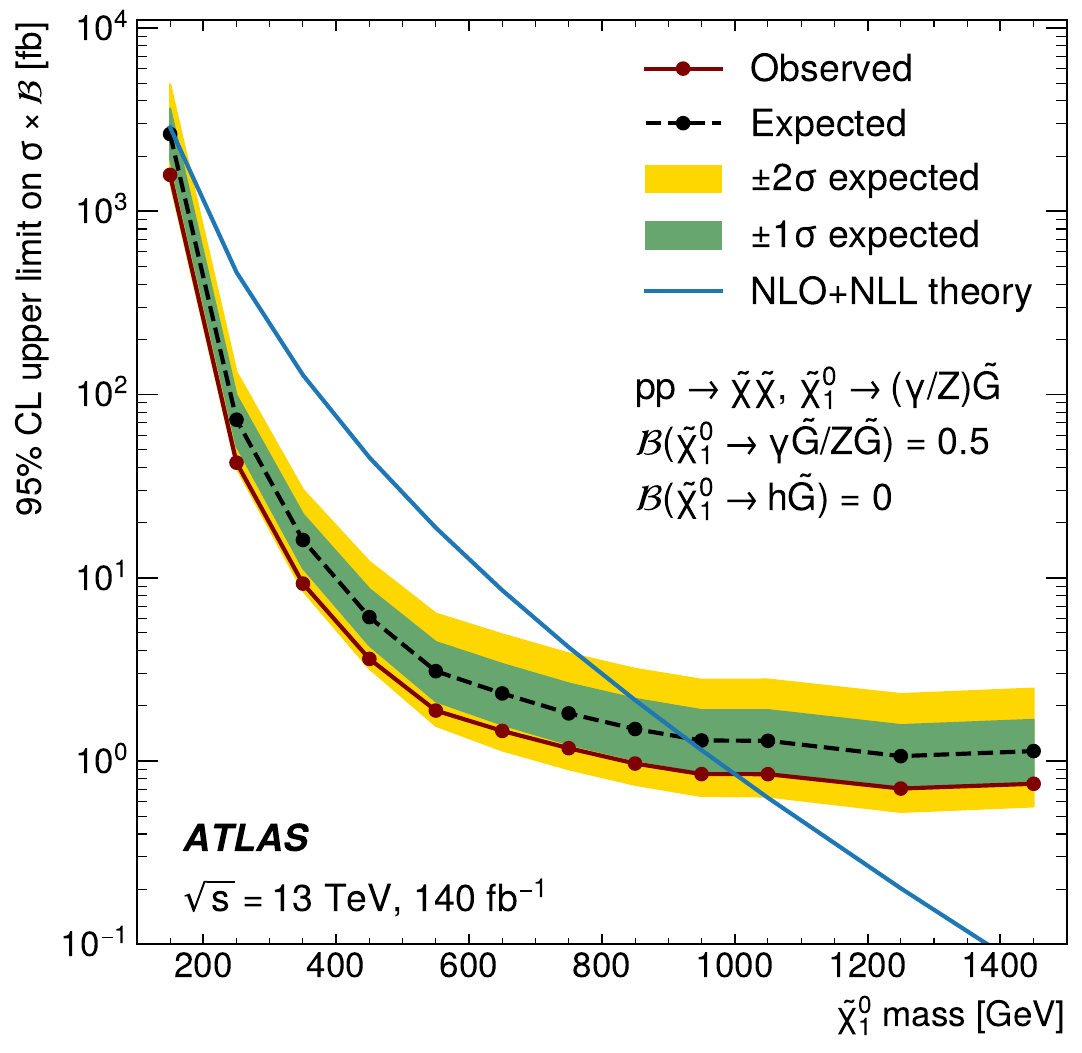}
\caption{}
\end{subfigure}
\begin{subfigure}[h]{0.49\linewidth}
\centering
\includegraphics[width=\textwidth]{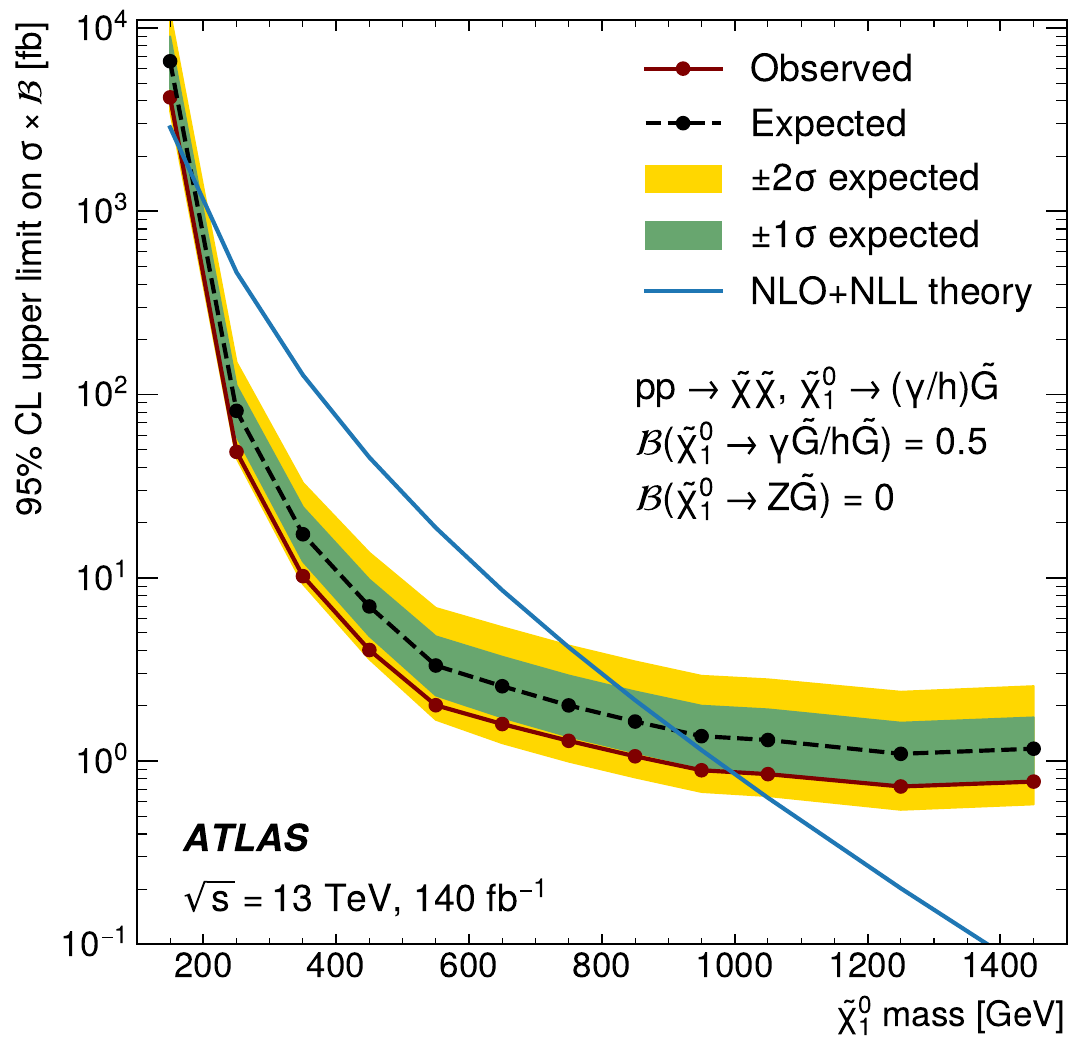}
\caption{}
\end{subfigure}
\caption{Observed and expected exclusion limits in the cross-section as a function of the \ninoone mass, at 95\% CL combining the three exclusion signal regions, for the full Run~2 data sample corresponding to an integrated luminosity of {\lumi}, for (a) the $\gam+Z$ and (b) the $\gam+h$ benchmark models.}
\label{fig:limit_XS}
\end{figure}

\begin{figure}
\centering
\begin{subfigure}[h]{0.49\linewidth}
\centering
\includegraphics[width=\textwidth]{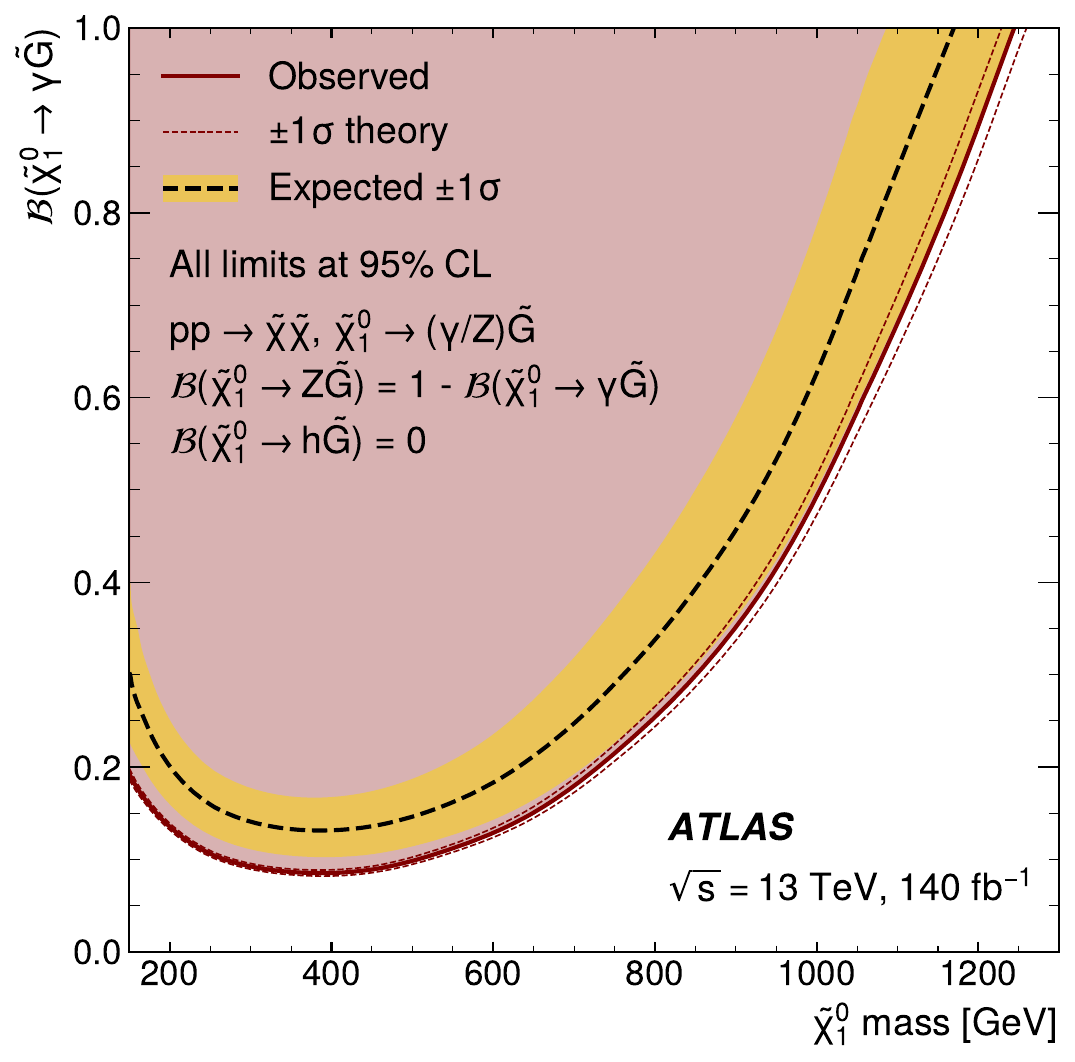}
\caption{}
\end{subfigure}
\begin{subfigure}[h]{0.49\linewidth}
\centering
\includegraphics[width=\textwidth]{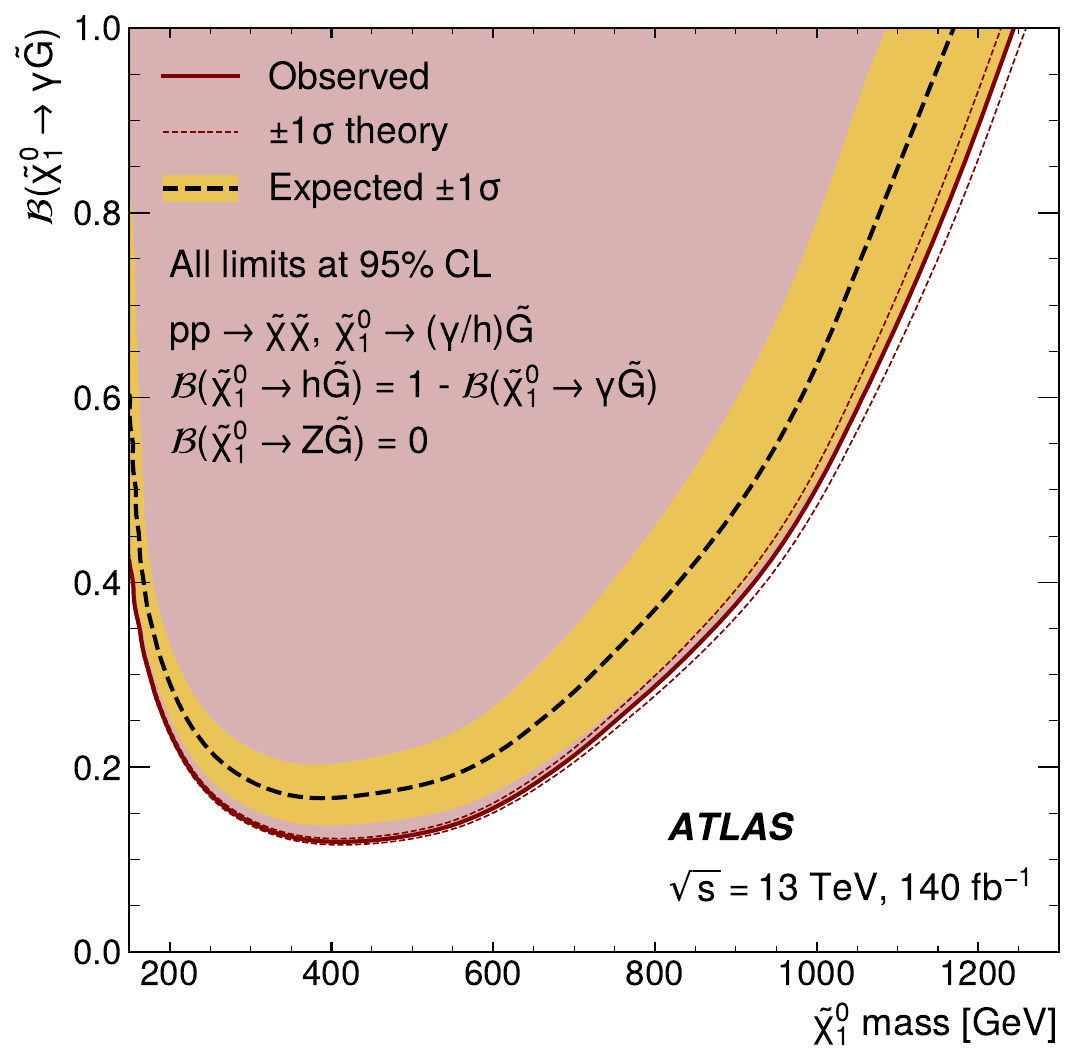}
\caption{}
\end{subfigure}
\caption{Observed and expected exclusion limits in the branching ratio to photons of the \ninoone as a function of its mass, at 95\% CL\@. In (a) the branching ratio to the Higgs boson is fixed to zero, while in (b) the branching ratio to the $Z$ boson is fixed to zero. The shaded area indicates excluded values.}
\label{fig:limit_BR}
\end{figure}

\begin{figure}
\centering
\begin{subfigure}[h]{0.49\linewidth}
\centering
\includegraphics[width=\textwidth]{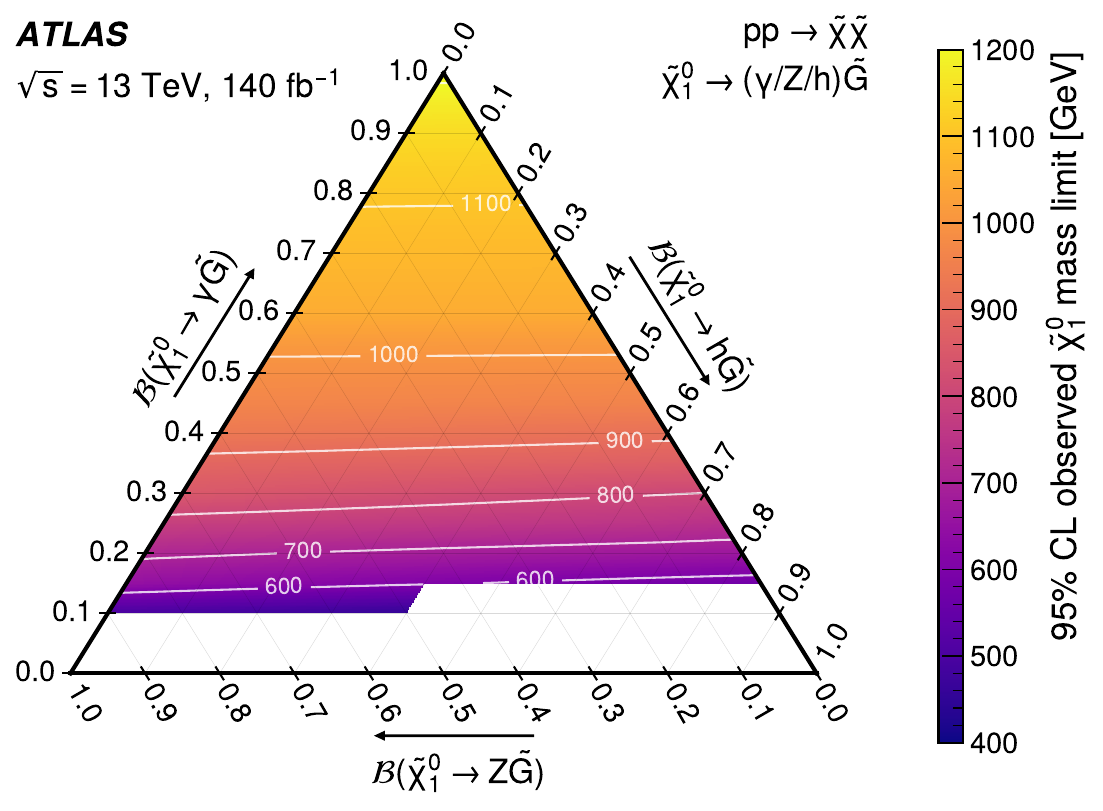}
\caption{}
\end{subfigure}
\begin{subfigure}[h]{0.49\linewidth}
\centering
\includegraphics[width=\textwidth]{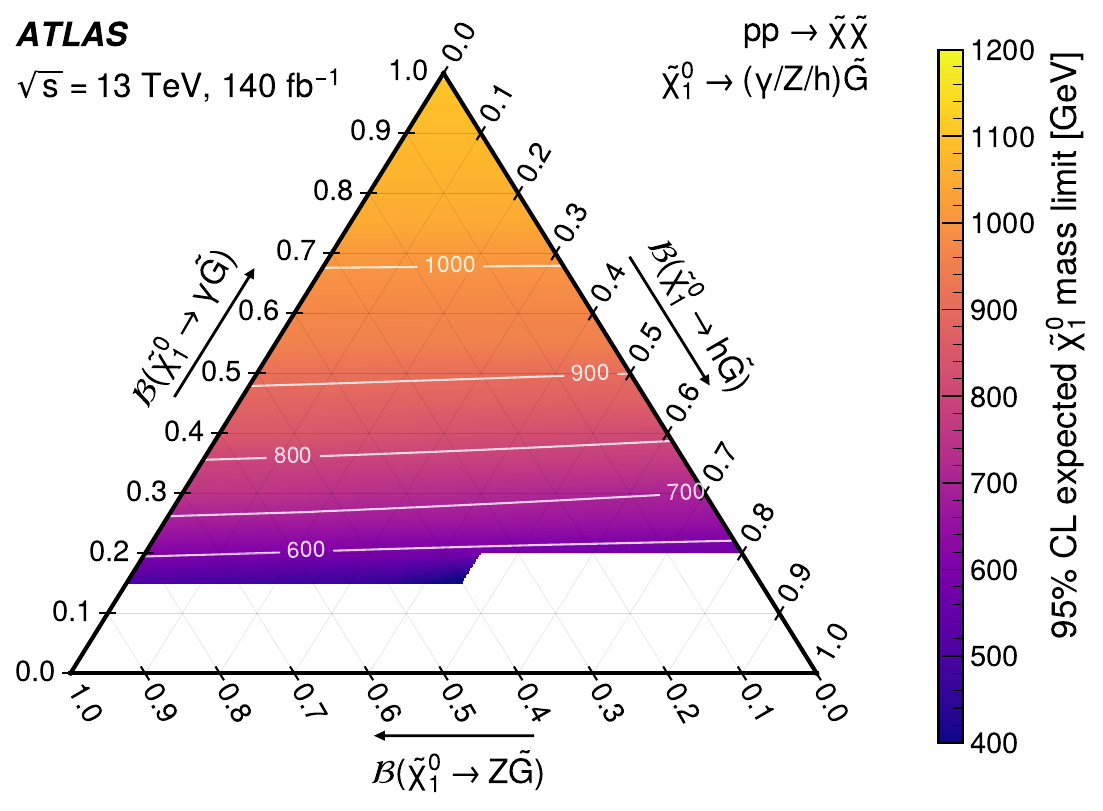}
\caption{}
\end{subfigure}
\caption{(a) Observed and (b) expected exclusion limits on the {\ninoone} mass at 95\% CL in the plane of branching ratios $\mathcal{B}$($\tilde{\chi}_1^0 \to \gamma \tilde{G}$), $\mathcal{B}$($\tilde{\chi}_1^0 \to Z \tilde{G}$), and $\mathcal{B}$($\tilde{\chi}_1^0 \to h \tilde{G}$) that satisfy $\mathcal{B}$($\tilde{\chi}_1^0 \to \gamma \tilde{G}$) + $\mathcal{B}$($\tilde{\chi}_1^0 \to Z \tilde{G}$) + $\mathcal{B}$($\tilde{\chi}_1^0 \to h \tilde{G}$) = 1. The colour scale indicates the excluded neutralino mass in \GeV, while the contour lines mark regions of constant excluded mass values.}
\label{fig:mass_limit_triangle}
\end{figure}

In view of the absence of an excess in the observed data relative to predictions the results are also used to establish exclusion limits in the various signal models. The signal contamination in the CRs and the experimental systematic uncertainties in the signal are taken into account for this calculation.
All the independent SRs (SRLe, SRMe and SRH) are statistically combined to set model-dependent upper limits on the cross-sections times branching ratios (Figure \ref{fig:limit_XS}), masses and variable branching ratios (Figures \ref{fig:limit_BR} and \ref{fig:mass_limit_triangle}).
The exclusion reach is highest in the pure photon branching ratio space, with a maximum around 1.2 {\TeV}, gradually decreasing to 450 {\GeV} for $\mathcal{B}(\ninoone\to\gamma\tilde{G}) = 10\%$ where the decays to $Z$ or $h$ bosons dominate. For those regions where the sensitivity is lower the low mass region can not be excluded, as can be seen in Figure \ref{fig:limit_BR}, and the lowest mass that can be excluded lies between 250 {\GeV} and 170 \GeV. The observed limits are stronger than the expected limits in accordance with the observed deficit in the data.

\clearpage


%
\section{Conclusion}
%
%

A search is presented for an excess of events featuring at least one photon, at least a jet and missing transverse momentum, as anticipated in scenarios where the lightest neutralino {\ninoone} is a mixture of bino and higgsino states. LHC proton--proton collisions at a centre-of-mass energy of \(\sqrt{s}=13\)~\TeV are analysed using data collected by the ATLAS detector during the complete 2015--2018 data-taking period, corresponding to a total integrated luminosity of \lumi.

Two strategies are employed: one optimised for discovery and another combining several non-overlapping signal regions in a simultaneous fit to achieve optimal exclusion limits for benchmark signals.

No excess is found above the predicted background in any of the signal regions. Model-independent limits are set on the visible cross-section for new physics processes. Exclusion limits are set on GGM models with LSP masses up to 1000~\GeV in two reference final-state scenarios containing a $Z$ or a Higgs boson in addition to the photon. The results are also interpreted in a model with variable branching ratios setting limits that range from 400~\GeV to 1200~\GeV.


\clearpage

\section*{Acknowledgements}
%

%
%

%
%

We thank CERN for the very successful operation of the LHC and its injectors, as well as the support staff at
CERN and at our institutions worldwide without whom ATLAS could not be operated efficiently.

The crucial computing support from all WLCG partners is acknowledged gratefully, in particular from CERN, the ATLAS Tier-1 facilities at TRIUMF/SFU (Canada), NDGF (Denmark, Norway, Sweden), CC-IN2P3 (France), KIT/GridKA (Germany), INFN-CNAF (Italy), NL-T1 (Netherlands), PIC (Spain), RAL (UK) and BNL (USA), the Tier-2 facilities worldwide and large non-WLCG resource providers. Major contributors of computing resources are listed in Ref.~\cite{ATL-SOFT-PUB-2025-001}.

We gratefully acknowledge the support of ANPCyT, Argentina; YerPhI, Armenia; ARC, Australia; BMWFW and FWF, Austria; ANAS, Azerbaijan; CNPq and FAPESP, Brazil; NSERC, NRC and CFI, Canada; CERN; ANID, Chile; CAS, MOST and NSFC, China; Minciencias, Colombia; MEYS CR, Czech Republic; DNRF and DNSRC, Denmark; IN2P3-CNRS and CEA-DRF/IRFU, France; SRNSFG, Georgia; BMFTR, HGF and MPG, Germany; GSRI, Greece; RGC and Hong Kong SAR, China; ICHEP and Academy of Sciences and Humanities, Israel; INFN, Italy; MEXT and JSPS, Japan; CNRST, Morocco; NWO, Netherlands; RCN, Norway; MNiSW, Poland; FCT, Portugal; MNE/IFA, Romania; MSTDI, Serbia; MSSR, Slovakia; ARIS and MVZI, Slovenia; DSI/NRF, South Africa; MICIU/AEI, Spain; SRC and Wallenberg Foundation, Sweden; SERI, SNSF and Cantons of Bern and Geneva, Switzerland; NSTC, Taipei; TENMAK, T\"urkiye; STFC/UKRI, United Kingdom; DOE and NSF, United States of America.

Individual groups and members have received support from BCKDF, CANARIE, CRC and DRAC, Canada; CERN-CZ, FORTE and PRIMUS, Czech Republic; COST, ERC, ERDF, Horizon 2020, ICSC-NextGenerationEU and Marie Sk{\l}odowska-Curie Actions, European Union; Investissements d'Avenir Labex, Investissements d'Avenir Idex and ANR, France; DFG and AvH Foundation, Germany; Herakleitos, Thales and Aristeia programmes co-financed by EU-ESF and the Greek NSRF, Greece; BSF-NSF and MINERVA, Israel; NCN and NAWA, Poland; La Caixa Banking Foundation, CERCA Programme Generalitat de Catalunya and PROMETEO and GenT Programmes Generalitat Valenciana, Spain; G\"{o}ran Gustafssons Stiftelse, Sweden; The Royal Society and Leverhulme Trust, United Kingdom.

In addition, individual members wish to acknowledge support from CERN: European Organization for Nuclear Research (CERN DOCT); Chile: Agencia Nacional de Investigaci\'on y Desarrollo (FONDECYT 1230812, FONDECYT 1240864, Fondecyt 3240661, Fondecyt Regular 1240721); China: Chinese Ministry of Science and Technology (MOST-2023YFA1605700, MOST-2023YFA1609300), National Natural Science Foundation of China (NSFC - 12175119, NSFC 12275265); Czech Republic: Czech Science Foundation (GACR - 24-11373S), Ministry of Education Youth and Sports (ERC-CZ-LL2327, FORTE CZ.02.01.01/00/22\_008/0004632), PRIMUS Research Programme (PRIMUS/21/SCI/017); EU: H2020 European Research Council (ERC - 101002463); European Union: European Research Council (BARD No. 101116429, ERC - 948254, ERC 101089007), European Regional Development Fund (SMASH COFUND 101081355, SLO ERDF), Horizon 2020 Framework Programme (MUCCA - CHIST-ERA-19-XAI-00), European Union, Future Artificial Intelligence Research (FAIR-NextGenerationEU PE00000013), Italian Center for High Performance Computing, Big Data and Quantum Computing (ICSC, NextGenerationEU); France: Agence Nationale de la Recherche (ANR-21-CE31-0022, ANR-22-EDIR-0002, ANR-24-CE31-0504-01); Germany: Baden-Württemberg Stiftung (BW Stiftung-Postdoc Eliteprogramme), Deutsche Forschungsgemeinschaft (DFG - 469666862, DFG - CR 312/5-2); China: Research Grants Council (GRF); Italy: Istituto Nazionale di Fisica Nucleare (ICSC, NextGenerationEU), Ministero dell'Università e della Ricerca (NextGenEU 153D23001490006 M4C2.1.1, NextGenEU I53D23000820006 M4C2.1.1, NextGenEU I53D23001490006 M4C2.1.1, SOE2024\_0000023); Japan: Japan Society for the Promotion of Science (JSPS KAKENHI JP22H01227, JSPS KAKENHI JP22H04944, JSPS KAKENHI JP22KK0227, JSPS KAKENHI JP24K23939, JSPS KAKENHI JP25H00650, JSPS KAKENHI JP25H01291, JSPS KAKENHI JP25K01023); Norway: Research Council of Norway (RCN-314472); Poland: Ministry of Science and Higher Education (IDUB AGH, POB8, D4 no 9722), Polish National Science Centre (NCN 2021/42/E/ST2/00350, NCN OPUS 2023/51/B/ST2/02507, NCN OPUS nr 2022/47/B/ST2/03059, NCN UMO-2019/34/E/ST2/00393, UMO-2022/47/O/ST2/00148, UMO-2023/49/B/ST2/04085, UMO-2023/51/B/ST2/00920, UMO-2024/53/N/ST2/00869); Portugal: Foundation for Science and Technology (FCT); Spain: Generalitat Valenciana (ASFAE/2022/008), Ministry of Science and Innovation (MCIN \& NextGenEU PCI2022-135018-2, MICIN \& FEDER PID2021-125273NB, RYC2019-028510-I, RYC2020-030254-I, RYC2021-031273-I, RYC2022-038164-I), Ministerio de Ciencia, Innovación y Universidades/Agencia Estatal de Investigaci\'on (PID2022-142604OB-C22); Sweden: Carl Trygger Foundation (Carl Trygger Foundation CTS 22:2312), Swedish Research Council (Swedish Research Council 2023-04654, VR 2021-03651, VR 2022-03845, VR 2022-04683, VR 2023-03403, VR 2024-05451), Knut and Alice Wallenberg Foundation (KAW 2018.0458, KAW 2022.0358, KAW 2023.0366); Switzerland: Swiss National Science Foundation (SNSF - PCEFP2\_194658); United Kingdom: Leverhulme Trust (Leverhulme Trust RPG-2020-004), Royal Society (NIF-R1-231091); United States of America: U.S. Department of Energy (ECA DE-AC02-76SF00515), Neubauer Family Foundation.

%
%


%
%
%
%
%
%
%

%
%
%
%
%
%

%
%

%
%
%
\printbibliography
\clearpage
 
\begin{flushleft}
\hypersetup{urlcolor=black}
{\Large The ATLAS Collaboration}

\bigskip

\AtlasOrcid[0000-0002-6665-4934]{G.~Aad}$^\textrm{\scriptsize 104}$,
\AtlasOrcid[0000-0001-7616-1554]{E.~Aakvaag}$^\textrm{\scriptsize 17}$,
\AtlasOrcid[0000-0002-5888-2734]{B.~Abbott}$^\textrm{\scriptsize 123}$,
\AtlasOrcid[0000-0002-0287-5869]{S.~Abdelhameed}$^\textrm{\scriptsize 119a}$,
\AtlasOrcid[0000-0002-1002-1652]{K.~Abeling}$^\textrm{\scriptsize 55}$,
\AtlasOrcid[0000-0001-5763-2760]{N.J.~Abicht}$^\textrm{\scriptsize 49}$,
\AtlasOrcid[0000-0002-8496-9294]{S.H.~Abidi}$^\textrm{\scriptsize 30}$,
\AtlasOrcid[0009-0003-6578-220X]{M.~Aboelela}$^\textrm{\scriptsize 45}$,
\AtlasOrcid[0000-0002-9987-2292]{A.~Aboulhorma}$^\textrm{\scriptsize 36e}$,
\AtlasOrcid[0000-0001-5329-6640]{H.~Abramowicz}$^\textrm{\scriptsize 157}$
% ...
\bigskip
\\

% ...
$^{17}$Department for Physics and Technology, University of Bergen, Bergen; Norway.\\
% ...
$^{30}$Physics Department, Brookhaven National Laboratory, Upton NY; United States of America.\\
% ...
$^{36}$$^{(a)}$Facult\'e des Sciences Ain Chock, Universit\'e Hassan II de Casablanca;$^{(b)}$Facult\'{e} des Sciences, Universit\'{e} Ibn-Tofail, K\'{e}nitra;$^{(c)}$Facult\'e des Sciences Semlalia, Universit\'e Cadi Ayyad, LPHEA-Marrakech;$^{(d)}$LPMR, Facult\'e des Sciences, Universit\'e Mohamed Premier, Oujda;$^{(e)}$Facult\'e des sciences, Universit\'e Mohammed V, Rabat;$^{(f)}$Institute of Applied Physics, Mohammed VI Polytechnic University, Ben Guerir; Morocco.\\
% ...
$^{45}$Physics Department, Southern Methodist University, Dallas TX; United States of America.\\
% ...
$^{49}$Fakult\"{a}t Physik, Technische Universit{\"a}t Dortmund, Dortmund; Germany.\\
% ...
$^{55}$II. Physikalisches Institut, Georg-August-Universit\"{a}t G\"ottingen, G\"ottingen; Germany.\\
% ...
$^{104}$CPPM, Aix-Marseille Universit\'e, CNRS/IN2P3, Marseille; France.\\
% ...
$^{119}$$^{(a)}$New York University Abu Dhabi, Abu Dhabi;$^{(b)}$United Arab Emirates University, Al Ain; United Arab Emirates.\\
% ...
$^{123}$Homer L. Dodge Department of Physics and Astronomy, University of Oklahoma, Norman OK; United States of America.\\
% ...
$^{157}$Raymond and Beverly Sackler School of Physics and Astronomy, Tel Aviv University, Tel Aviv; Israel.\\
% ...
$^{a}$ Also at Affiliated with an institute formerly covered by a cooperation agreement with CERN.\\
$^{b}$ Also at An-Najah National University, Nablus; Palestine.\\
$^{c}$ Also at Borough of Manhattan Community College, City University of New York, New York NY; United States of America.\\
$^{d}$ Also at Center for Interdisciplinary Research and Innovation (CIRI-AUTH), Thessaloniki; Greece.\\
$^{e}$ Also at Centre of Physics of the Universities of Minho and Porto (CF-UM-UP); Portugal.\\
$^{f}$ Also at CERN, Geneva; Switzerland.\\
% ...

\end{flushleft}


%

%
%
\clearpage

\end{document}